\documentclass[%
 reprint,
superscriptaddress,
 amsmath,amssymb,
 aps,
prm,
]{revtex4-2}
\usepackage{xcolor}

\usepackage{graphicx}
\usepackage{epstopdf}
\usepackage{dcolumn}
\usepackage{bm}
\usepackage{gensymb}
\usepackage{amsfonts}
\usepackage{upgreek}
\usepackage{hyperref}
\usepackage{multirow}
\usepackage{tabularx}
\usepackage[section]{placeins}
\footnotetext{These authors contributed equally to this work.}
\begin{document}

\author{Frank M. Abel$^\dagger$}
\email{frank.abel@nist.gov}
\affiliation{Physics Department, United States Naval Academy, Annapolis, MD 21402, USA}
\affiliation{National Institute of Standards and Technology, Gaithersburg, MD 20899, USA}

\author{Subhash Bhatt$^\dagger$}
\affiliation{Department of Physics and Astronomy, University of Delaware, Newark, DE 19716, USA}

\author{Shelby S. Fields$^\dagger$}
\affiliation{Materials Science and Technology Division, U.S. Naval Research Laboratory, Washington, District of Columbia 20375, USA}

\author{Vinay Sharma}
\affiliation{Laboratory for Physical Sciences, College Park, MD 20740, USA}

\author{Dai Q. Ho}
\affiliation{Department of Materials Science and Engineering, University of Delaware, Newark, Delaware 19716, USA}
\affiliation{Faculty of Natural Sciences, Quy Nhon University, Quy Nhon 55113, Vietnam}

\author{Daniel Wines}
\affiliation{National Institute of Standards and Technology, Gaithersburg, MD 20899, USA}

\author{D. Quang To}
\affiliation{Department of Materials Science and Engineering, University of Delaware, Newark, Delaware 19716, USA}

\author{Joseph C. Prestigiacomo}
\affiliation{Materials Science and Technology Division, U.S. Naval Research Laboratory, Washington, District of Columbia 20375, USA}

\author{Tehseen Adel}
\affiliation{National Institute of Standards and Technology, Gaithersburg, MD 20899, USA}
\affiliation{Department of Physical Sciences, University of Findlay, Findlay, OH 45840}

\author{Riccardo Torsi}
\affiliation{National Institute of Standards and Technology, Gaithersburg, MD 20899, USA}

\author{Maria F. Munoz}
\affiliation{National Institute of Standards and Technology, Gaithersburg, MD 20899, USA}

\author{David T. Plouff}
\affiliation{Department of Physics and Astronomy, University of Delaware, Newark, DE 19716, USA}

\author{Xinhao Wang}
\affiliation{Department of Physics and Astronomy, University of Delaware, Newark, DE 19716, USA}

\author{Brian Donovan}
\affiliation{Physics Department, United States Naval Academy, Annapolis, MD 21402, USA}

\author{Don Heiman}
\affiliation{Department of Physics, Northeastern University, Boston, MA 02115, USA}
\affiliation{Plasma Science and Fusion Center, MIT, Cambridge, MA 02139, USA}

\author{Gregory M. Stephen}
\affiliation{Laboratory for Physical Sciences, College Park, MD 20740, USA}

\author{Adam L. Friedman}
\affiliation{Laboratory for Physical Sciences, College Park, MD 20740, USA}

\author{Garnett W. Bryant}
\affiliation{National Institute of Standards and Technology, Gaithersburg, MD 20899, USA}
\affiliation{Joint Quantum Institute, University of Maryland, College Park, Maryland 20742, USA}

\author{Anderson Janotti}
\affiliation{Department of Materials Science and Engineering, University of Delaware, Newark, Delaware 19716, USA}

\author{Michelle E. Jamer}
\affiliation{Physics Department, United States Naval Academy, Annapolis, MD 21402, USA}

\author{Angela R. Hight Walker}
\affiliation{National Institute of Standards and Technology, Gaithersburg, MD 20899, USA}

\author{John Q. Xiao}
\affiliation{Department of Physics and Astronomy, University of Delaware, Newark, DE 19716, USA}

\author{Steven P. Bennett}
\affiliation{Materials Science and Technology Division, U.S. Naval Research Laboratory, Washington, District of Columbia 20375, USA}

\date{\today}

\begin{abstract}
RuO$_{2}$ has been proposed as the prototypical altermagnetic material. However, several reports have recently questioned its intrinsic magnetic ordering, leading to conflicting findings, especially in thin film heterostructures pointing to possible interface effects being convoluted with supposed antiferromagnetic/altermagnetic signatures. Here, extensive magnetometry measurements were performed on two independently grown thin film heterostructures of RuO$_{2}$ interfaced with either NiFe or Fe acting as the ferromagnetic layer. Below about 15 K, both samples exhibit exchange bias fields when cooled to approximately 2 K in a $+$1 T field, and a spin transitional feature is observed around 31 K. Magneto-Raman measurements on RuO$_{2}$ thin films only reveal a magnon mode when there is a NiFe layer, suggesting that RuO$_{2}$ does not intrinsically possess long range magnetic ordering.. When in contact with a ferromagnet, RuO$_2$ displays effects that could be ascribed to antiferromagnetism. However, the lack of intrinsic magnon modes points toward possible diffusion between the layers or spin disorder at the interface as seen by density functional theory (DFT) calculations.
\end{abstract}

\title{Probing Magnetic Properties of RuO$_{2}$ Heterostructures Through the Ferromagnetic Layer}

\maketitle

\section*{Introduction}

Altermagnets are an exciting new class of magnetic materials, where opposite spin-sublattices in a collinear antiferromagnet are connected by a rotation symmetry, which results in spin-split energy bands, enabling new possibilities for applications.\cite{PhysRevX.12.040501} In parallel, there has recently been a large push to develop antiferromagnetic (AFM) materials for use in spintronics to overcome the limitations in ferromagnetic (FM)-based devices including stray fields, lower switching speed, and size constraints from domain structures.\cite{jungfleisch2018perspectives} Altermagnets are a proposed solution due to their net-zero magnetization and band splitting which enable ferromagnetic properties along with faster switching speeds in the THz range.\cite{10.1002/adfm.202313332, PhysRevApplied.21.034038} 
\par
Historically, RuO$_{2}$ was classified as a Pauli paramagnet \cite{ryden1970magnetic, Fletcher1968,over2012surface,PhysRevB.60.12279}. In 2017 neutron diffraction measurements suggested it may be an itinerant antiferromagnet with an above room temperature N\'eel temperature. \cite{berlijn2017itinerant} Resonate X-ray scattering and X-ray magnetic linear dichroism also showed signatures of collinear AFM state.\cite{PhysRevLett.122.017202,yi2025probing} Recently, strained RuO$_2$ thin films have been shown to exhibit superconductivity, leading to more interest in the material.\cite{PhysRevLett.125.147001,10.1038/s41467-020-20252-7} With the advent of altermagnetic theory, RuO$_{2}$ has been proposed as one of the prototypical materials with studies observing time reversal symmetry breaking\cite{doi:10.1126/sciadv.adj4883} and the anomalous Hall effect (AHE). \cite{Liu2022, tschirner2023saturation} However, in the past year, there has been an influx of studies questioning the intrinsic magnetic ordering in this material. \cite{PhysRevB.109.134424, PhysRevLett.132.166702, npjSpintronics2024, PhysRevLett.133.176401,plouff2025revisiting} In bulk crystals, muon spin rotation showed that there was no evidence of a magnetic phase between 5 K and 400 K, estimating an upper limit for the moment to be $4.8 \times 10^{-4}\,\mu_\mathrm{B}$/Ru.\cite{PhysRevLett.132.166702} Another study by Ke{\ss}ler {\it{et al.}} using both neutron diffraction and muon spectroscopy showed that both bulk and thin films lacked evidence for magnetic ordering,\cite{npjSpintronics2024} leading to the conclusion that the single crystals are not magnetic.\cite{kiefer2025crystal} Using a combination of muon spectroscopy and density functional theory (DFT) the maximum moment was estimated to be $1.4 \times 10^{-4}\,\mu_\mathrm{B}$/Ru and $7.5 \times 10^{-4}\,\mu_\mathrm{B}$/Ru, for bulk crystals and epitaxial thin films, respectively.\cite{PhysRevLett.132.166702,npjSpintronics2024} Additionally, polarized neutron reflectometry (PNR) at 300 K showed no net magnetization for uniform  (101) oriented RuO$_{2}$ films grown on TiO$_{2}$ in either the bulk or interface of the film.\cite{wu2025magnetic} There has been controversy in the spin-ARPES results where J. Liu {\it{et al.}} demonstrated a lack of spin splitting by spin- and angle-resolved photoemission spectroscopy (spin-ARPES) in both thin films and bulk and is further supported by theoretical results,\cite{PhysRevLett.133.176401,osumi2025spindegeneratebulkbandstopological} which contradicts a report on the observation of bulk band structure in single crystals.\cite{PhysRevB.111.134450} Despite the uncertainty over the altermagnetic state of RuO$_2$, reports still claim evidence for altermagnetic and antiferromagnetic signatures particularly in thin film heterostructures and device stacks.\cite{10.1002/adfm.202313332, liao2024separation, PhysRevApplied.21.034038, 10.1063/5.0213320,bose2022tilted,ahn2019antiferromagnetism,bai2023efficient} These contradictory results have only fueled interest in further investigating this material for possible magnetic ordering and methods to probe the magnetic and possible altermagnetic ordering.\cite{PhysRevB.109.094413,PhysRevB.108.L100402,PhysRevB.110.L100402,wenzel2025fermi}
\par
Recently, two studies seem to suggest that interfacial effects are a driving mechanism for many of the observed phenomenon in thin film heterostructures. In SrRuO$_{3}$/RuO$_{2}$ heterostructure an exchange bias (EB) effect is observed and an enhanced AHE is measured in the heterostructure suggesting a possible non-zero magnetic moment in the RuO$_{2}$ layer. \cite{10.1002/smll.202408247} A transport study of a RuO$_{2}$/NiFe shows the possibility of an interface generated spin current which needs to be accounted for when claiming altermagnetic effects. \cite{akashdeep2025interface} Its also has been shown that in very thin (1.7 nm) (110) RuO$_2$, sandwiched between TiO$_2$, exhibits AHE at lower magnetic fields then previously reported, showing nonlinear behavior in the Hall resistivity vs. magnetic field between 1.8 K and 15 K. This study suggests a strain or surface effects being strongly related to possible magnetism in RuO$_2$. \cite{jeong2025metallicity}
\par
In our study, we extensively investigate two, independently grown RuO$_2$ heterostructures on different single-crystal substrates (MgF$_{2}$ and TiO$_{2}$) topped with two different ferromagnetic layers, NiFe and Fe. Both samples exhibit an EB field below 15 K and a spin-like transition feature at approximately 31 K. Temperature-dependent magneto-Raman is used to probe long-range magnetic ordering and interfacial effects by searching for new magnon modes in bare RuO$_2$ films and a ferromagnetic heterostructure. Our results suggest that RuO$_2$ does not possess intrinsic magnon modes or induced interfacial magnon modes, indicating that the emergence of the EB likely comes from frustrated spins at the interface, potentially arising from a variety of effects. Density functional theory (DFT) calculations of a RuO$_2$/Fe interface suggest that spin disorder can arise intrinsically, and that there can be weak AFM-like ordering present within the first few layers of the RuO$_2$. These results indicate that interfacial effects in RuO$_2$/FM heterostructures likely produce the EB and spin-like transition effect, which could be mistaken for signatures of AFM ordering or altermagnetism.

\section*{Structural Characterization}

The thin films in this study were grown by magnetron sputtering on either single-crystal TiO$_{2}$ or MgF$_{2}$ substrates, with details given in the methods section. The MgF$_{2}$ was primarily chosen due to having minimal phonon mode activity for the magneto-Raman spectroscopy study. Orientation of the RuO$_{2}$ was confirmed by X-ray diffraction (XRD), and corresponded to the same orientation of the single crystal substrate on which the film was grown. The two samples used in the primary magnetometry study were extensively evaluated by reciprocal space mapping (RSM), and the thicknesses determined by X-ray reflectivity (XRR), all other given thicknesses are based on deposition time. RSM measurements of asymmetric diffraction peaks collected for both heterostructures with (110) RuO$_{2}$ crystallographic orientation are shown in Fig.~\ref{fig:RSM}(a-d), the high-resolution(HR) XRD and XRR measurements of these samples are provided in the supplemental material, Fig. ~S1 and Fig. ~S2, respectively. Qualitatively, RSM maps of both heterostructures with different in-plane angular components boast RuO$_2$ peaks with confined intensity, consistent with high crystallinity within the rutile layer. The most direct comparison between the crystalline quality of each film is made through examination of peak sizes, where the (310) and (231) peaks from the MgF$_2$ heterostructure display narrower $\it{q}_x$ widths than their counterparts within the TiO$_2$ heterostructure, which is consistent with larger crystalline domains in this sample. In bulk RuO$_2$, the (310) peak is found at $\it{q}_x$ = 3.12 \AA, $\it{q}_z$ = 6.23 \AA. Whereas, (310) peaks from RuO$_2$ film layers within these heterostructures arise at $\it{q}_x$ = 3.03 \AA, $\it{q}_z$ = 6.09 {\AA} and $\it{q}_x$ = 3.04 \AA, $\it{q}_z$ = 6.08 \AA, respectively for MgF$_2$ and TiO$_2$. Accordingly, along the out-of-plane [110] and in-plane [1$\bar{1}$0] directions, both films experience tensile strain. Rotated about $25^\circ$ in-plane from the (310) peak, the (321)/(231) peak in bulk RuO$_2$ is found at $\it{q}$$_x$ = 3.54 \AA, $\it{q}$$_z$ = 7.78 \AA. In agreement with analysis of the (310) peak positions, the locations of the (321)/(231) RuO$_2$ film peaks in the MgF$_2$ ($\it{q}_x$ = 3.52 \AA, $\it{q}_z$ = 7.59 \AA) and TiO$_2$ ($\it{q}_x$ = 3.55 \AA, $\it{q}$$_z$ = 7.71) \AA) heterostructures  are indicative of tensile strain. It is evident, based on comparison of the signs of $\it{d}$-spacing shifts with the behaviors predicted through rigid body calculation (Table~\ref{tab:structure}), that the strain states of these heterostructures are not completely biaxial, and that additional factors such as the presence of microstructure, coefficient of thermal expansion mismatches, or strains induced by further processing are affecting the final strain states of the RuO$_2$ within the heterostructures.

\begin{figure*}[]
    \centering
    \includegraphics[width = 0.8\textwidth]{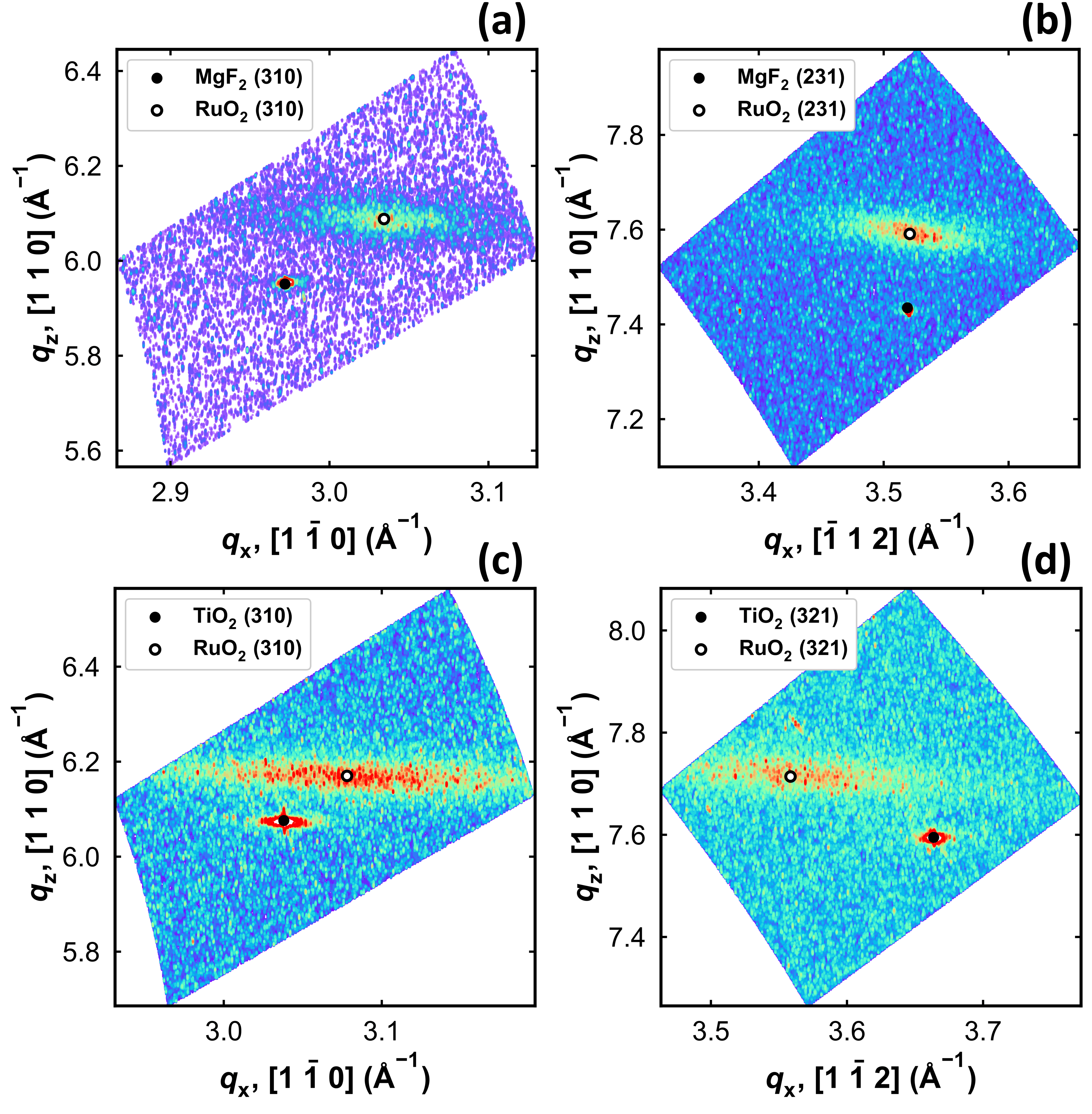}
     \caption{Reciprocal space mapping (RSM) of (a) (310) and (b) (231) film and substrate diffraction peaks from RuO$_2$ grown on MgF$_2$ and (c) (310) and (d) (321) film and substrate diffraction peaks from RuO$_2$ grown on TiO$_2$. Each plot is annotated with the nominal positions of the peaks, as indicated by each respective legend.}
    \label{fig:RSM}
\end{figure*}

\begin{table}[]
    \centering
    \begin{tabular}{|c|c|c|c|}
    \hline
     &  Layer  & MgF$_2$ &TiO$_2$ \\
\hline
 \multirow{3}{*}{Thickness (nm)}    & RuO$_2$ & 32.3 & 35.8 \\
    & FM &  8.8 & 9.5 \\
    & Capping Layer & 4.2 & 3.8 \\
    \hline
 \multirow{3}{*}{Lattice Constants}    & a (\AA) &  4.69 & 4.59 \\
    & c (\AA) &  3.09 & 2.95 \\
    & Strain (\%) & -4.5 & 1.4 \\
    \hline    
    \end{tabular}
    \caption{Thicknesses and structural properties of employed substrates. The values in the last row correspond to the calculated out-of-plane strains in the (110) direction in fully coherent RuO$_2$. The parameters for thickness are calculated from the XRR fit in the supporting material Fig. ~S2. The table shows the comparable nature of the samples chosen for the study including the thicknesses.}
    \label{tab:structure}
\end{table}

\section*{Magnetometry, Magneto-Raman, and DFT Calculations of RuO$_{2}$ Heterostructures}

When coupled with an FM material, the interfacial magnetic properties of RuO$_2$ can be probed through exchange interactions. EB canonically occurs when an AFM ordered magnet couples to a FM layer, leading to a shift (spin pinning) in the hysteresis loop when cooled through the N\'eel temperature of the antiferromagnet.\cite{NOGUES1999203,doi:10.1021/acsanm.8b02319} However, EB can also be observed in spin glass or spin disordered materials coupled to an FM, suggesting a true AFM material is not required to observe these effects.\cite{ali2007exchange} EB can indicate a surface or frustrated ordering at the surface of a secondary layer. Accordingly, the prepared heterostructures comprising MgF$_2$/RuO$_2$/NiFe/Al and TiO$_2$/RuO$_2$/Fe/Ru enable the exploration of the interfacial spin effects in RuO$_2$ by probing the presence and nature of exchange interactions with a ferromagnetic layer. The two samples explored in the main text were chosen due to their comparable nature in the thickness of the RuO$_2$ and ferromagnetic layer as noted in Table \ref{tab:structure}. In both sample sets, the EB was induced by heating the thin film to 400 K, then applying a $\pm$1 T field for 1 hour. Then, the samples were subsequently cooled to about 2 K (base temperature) in the 1 T field. The biasing temperature of 400 K was chosen due to practical limitations of the magnetometers, and due to the proposed N\'eel temperature being above about 300 K. \cite{berlijn2017itinerant, Liu2022}  
\par
\begin{figure*}[]
    \centering
    \includegraphics[width =0.95\textwidth]{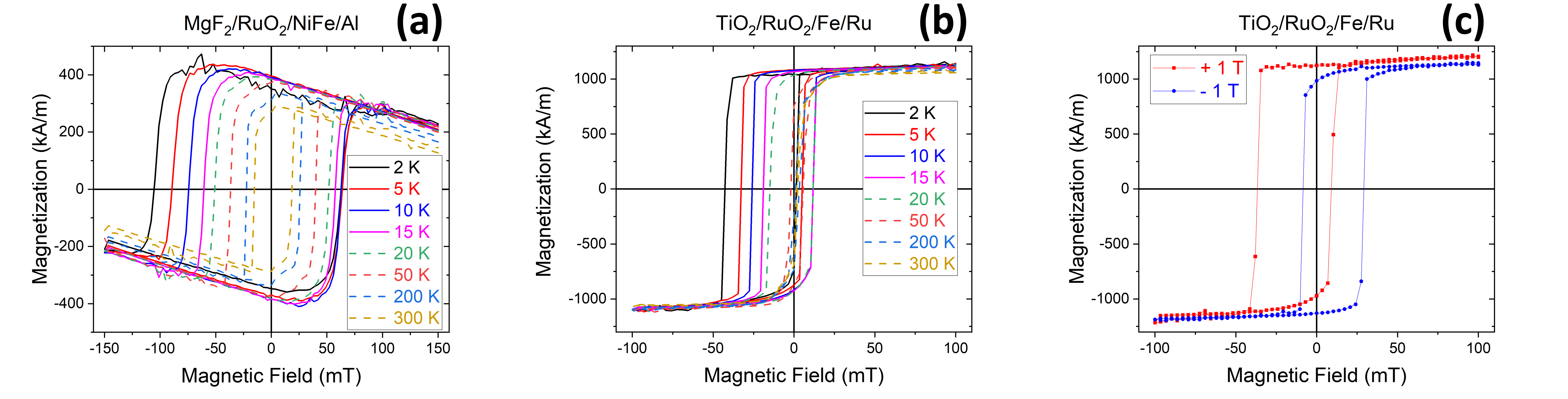}
    \caption{Select hysteresis loops measured on (a) MgF$_{2}$/RuO$_{2}$/NiFe/Al and (b) TiO$_{2}$/RuO$_{2}$/Fe/Ru on warming between 2 K and 300 K after biasing and field cooling from 400 K in $+$1T field. (c) Hysteresis loop measured at 5 K after biasing with positive (red) and negative (blue) 1 T field. The data is normalized by the volume of the ferromagnetic layer.}
    \label{fig:MH_EB}
\end{figure*}
\begin{figure*}[]
    \centering
    \includegraphics[width =0.95\textwidth]{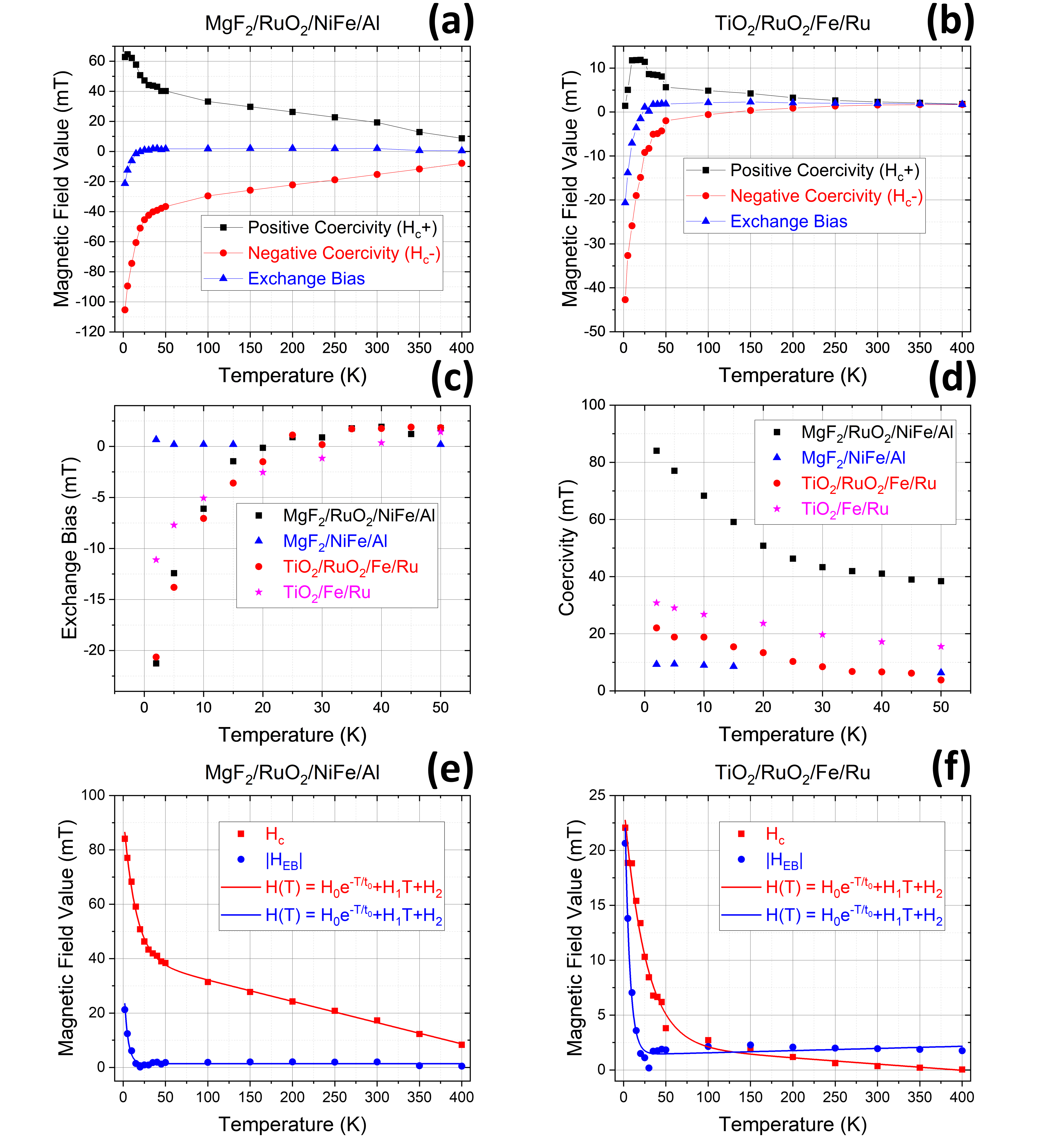}
    \caption{Positive coercivity (black squares), negative coercivity (red circles), and exchange bias (blue triangles) measured on (a) MgF$_{2}$/RuO$_{2}$/NiFe/Al and (b) TiO$_{2}$/RuO$_{2}$/Fe/Ru on warming from 2 K to 400 K after training and field cooling. Comparison of (c) exchange bias and (d) coercivity values measured  between 2 K and 50 K following training and field cooling of MgF$_{2}$/RuO$_{2}$/NiFe/Al (black squares) and TiO$_{2}$/RuO$_{2}$/Fe/Ru (red circles) heterostructures and MgF$_{2}$/RuO$_{2}$/NiFe/Al (blue triangles) and TiO$_{2}$/RuO$_{2}$/Fe/Ru (pink stars) control samples. Exponential fits of coercivity and absolute value of exchange bias (EB) field measured on (e) MgF$_{2}$/RuO$_{2}$/NiFe/Al and (f) TiO$_{2}$/RuO$_{2}$/Fe/Ru.}
    \label{fig:Temp_EB}
\end{figure*}

The magnetization (M) vs. magnetic field (H) loops in Fig.~\ref{fig:MH_EB}(a,b,c) were collected from  about 2 K to 300 K to keep the same EB setting condition. The hysteresis loops in the biased state have a large negative coercivity as shown in Fig.~\ref{fig:MH_EB}(a) and (b) for the MgF$_2$ and TiO$_2$ samples, respectively. In the MgF$_2$ sample, the hysteresis loop was larger than the TiO$_2$, most likely due to sample roughness leading to more pinned domains, as qualitatively assessed by XRR shown in the supplemental material, Table S~1. Fig.~\ref{fig:MH_EB}(c) shows measurements of the TiO$_2$ sample bias with both a positive and negative applied field, demonstrating a negative shift with a positive bias field and a positive shift with a negative bias field. However, these EB measurements illustrate that the sample is not fully biased in either direction. The positive coercivity ($H_c^+$), negative coercivity ($H_c^-$), and EB for both samples are shown in Fig.~\ref{fig:Temp_EB}. The EB and coercivity are calculated by using $H_{EB} = \frac{H_c^+ + H_c^-}{2}$ and $H_{c} = \frac{H_c^+ - H_c^-}{2}$, where $H_c^+$ and $H_c^-$ is the field where the magnetization is zero along the positive and negative axis, respectively. The asymmetry in the hysteresis loops is highlighted with the difference between the positive and negative coercivity, Fig.~\ref{fig:Temp_EB}(a,b). There is a similar exponential increase between the two sample sets in the negative coercivity from low temperature until the bias is extinguished. Examining the positive coercivity, the TiO$_2$ sample shows a pronounced step-like feature between about 2 K and 50 K, showing a maximum around 12 K. The feature is less apparent in the MgF$_2$ sample with a negligible change between 2 and 5 K, followed by a decrease with increasing temperature. Similar increase and decrease in coercivity have been noted in a spin glass ferromagnetic heterostructures.\cite{ali2007exchange}. Examination of the EB shows that both samples reach nearly identical value of -21 mT at 2 K and vanishes by 15 K. Fig.~\ref{fig:Temp_EB} (c,d) shows the EB and coercivity between 2 K and 50 K with comparison with two control samples of MgF$_2$/NiFe/Al and TiO$_2$/Fe/Ru. The MgF$_2$ control sample shows that there is no EB generated from only the NiFe ferromagnetic layer. However, the TiO$_2$/Fe/Ru shows a small EB that emerges over a similar temperature range, reaching a maximum of -11 mT at 2 K. The small EB suggests a portion of the observed EB effect could come from Fe-O bonding at the TiO$_{2}$ interface, diffusion at the Fe/Ru interface, or possible oxidation of the Ru capping layer leading to a more disordered Fe/Ru-O interface. In Fig.~\ref{fig:Temp_EB}(e,f), the coercive field ($H_c$) and absolute value of the EB versus temperature for both samples are shown. We observe a clear exponential decrease in both the value of $H_c$ and $H_{EB}$ as a function of temperature. The data are fit to an exponential function with a linear component \cite{10.1063/5.0202812} and the fitting parameters are reported in the supplemental material, Table ~S2. The exponential decrease in H$_c$ and H$_{EB}$ may give insight into the magnetic interactions present in the RuO$_2$ layer or at the interface when in contact with a ferromagnet as indicated by previous studies.\cite{nayak2020exchange, 10.1063/1.4941795, ding2013interfacial, moutis2001exchange} Examination of the EB shows that it is fully extinguished by about 15 K, as previously noted, and $H_c$ becomes purely linear at about 50 K for both samples. The exponential decay of the coercivity below 50 K is of particular note and similar coercivity behavior has been observed in Mn-based heuslers with AFM ordering and spin glass behavior\cite{10.1063/5.0202812} and AFM/FM heterostructures. \cite{nayak2020exchange, moutis2001exchange} This kind of behavior is often associated with magnetically frustrated systems where competition between magnetic domains can form a spin glass-like state. \cite{10.1063/1.4941795, karmakar2008evidence, huang2008size, ding2013interfacial} The findings suggest there is likely magnetic frustration present in the heterostructures, and the emergence of EB and coercivity behavior may be attributed to a spin glass-like state forming at the interface.
\par
\begin{figure*}[]
    \centering
    \includegraphics[width =0.95\textwidth]{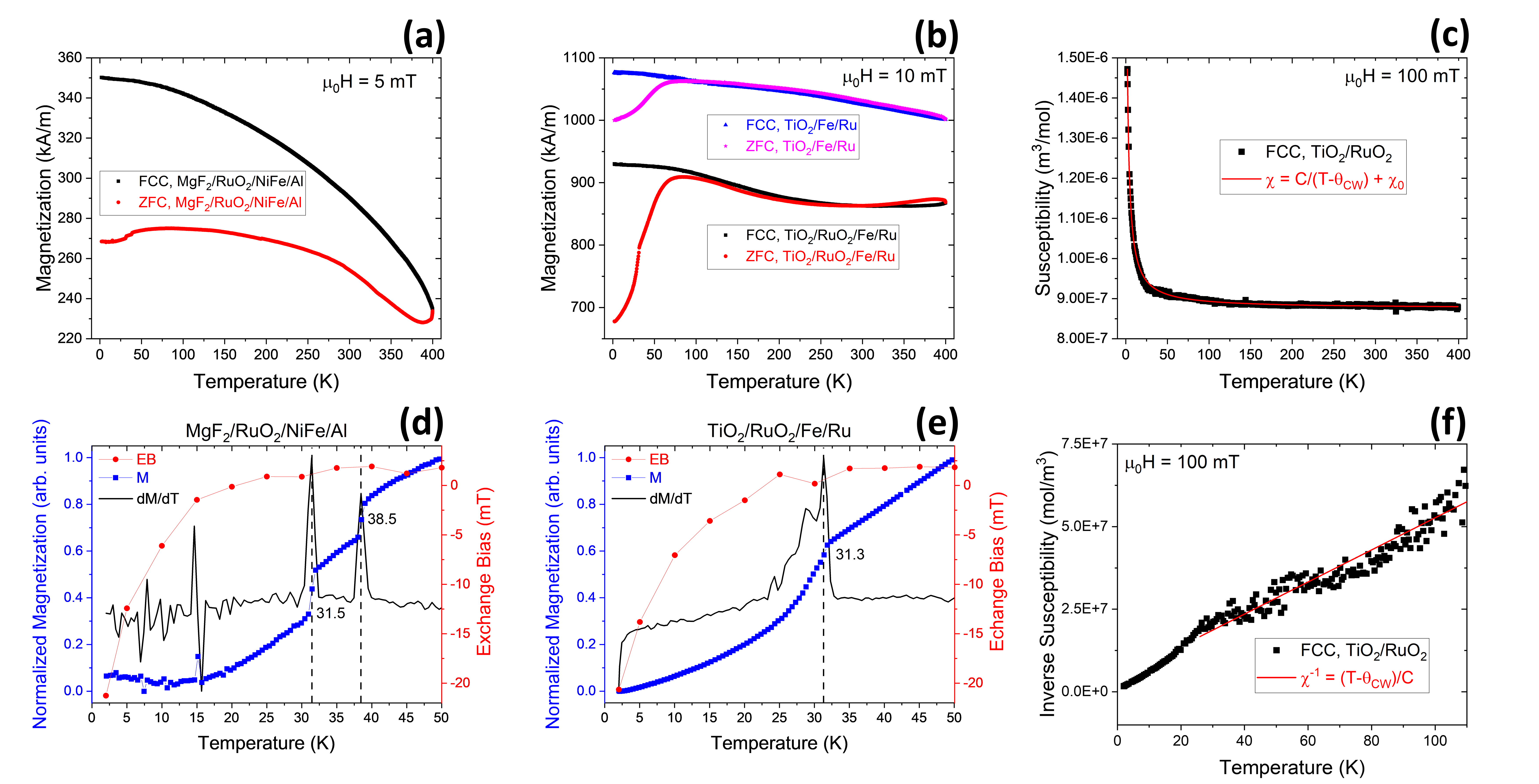}
    \caption{Zero-Field-Cooled (ZFC) and field-cooled-cooling (FCC) of MgF$_{2}$/RuO$_{2}$/NiFe/Al, TiO$_{2}$/RuO$_{2}$/Fe/Ru, and TiO2$_{2}$/Fe/Ru (a,b). Susceptibility determined from FCC measurement of TiO$_{2}$/RuO$_{2}$ fit to the Curie-Weiss law with constant component ($\chi_0$) (c). ZFC, exchange bias, and derivative of ZFC curves of MgF$_{2}$/RuO$_{2}$/NiFe/Al and TiO2$_{2}$/RuO$_{2}$/Fe/Ru highlighting spin-like transitions around 31 K (d,e). Inverse Susceptibility with constant $\chi_0$ from subtracted determined from FCC measurement of TiO$_{2}$/RuO$_{2}$ fit to the Curie-Weiss law (f). The data is normalized by the volume of the ferromagnetic layer for the heterostructure samples and the volume of the RuO$_{2}$ for the single layer sample.}
    \label{fig:ZFCFC}
\end{figure*}

The zero-field-cooled (ZFC) magnetization versus temperature measurements in Fig.~\ref{fig:ZFCFC}(a) and (b) and highlighted in (d) and (e) show a spin-like transition at about 31 K for both samples. The ZFC is also shown for the TiO$_{2}$ reference samples in (b) and notably lacks this feature despite showing a small EB. For the MgF$_2$ sample, there are two steps in the ZFC measurement in (d), which may be related to the roughness of the sample or the presence of both Ni/Fe, if these effects originate from an interface interaction. The less rough sample on TiO$_2$ in (e) has a cleaner transition at 31.3 K where there is a distinct change in the linearity of the magnetization measurement. It is evident, upon comparison of the EB behavior for each heterostructure, that EB is only present below the spin-like transition temperature. Further, ZFC curve derivatives in Fig.~\ref{fig:ZFCFC} (d, e), emphasize the sharpness of the transition. Additionally, the spin-like transition temperature coincides with the region of exponential decay observed in the coercivity which becomes linear only above about 50 K. In the case of the MgF$_2$ sample in Fig.~\ref{fig:ZFCFC}(a), we notice that the ZFC and field-cooled-cooling (FCC) curves do not converge similarly to the TiO$_2$ sample in Fig.~\ref{fig:ZFCFC}(b). The larger coercivity in the MgF$_2$ sample most likely leads to this irreversible magnetization behavior. 
The susceptibility, FCC at 100 mT, of single layer (56 nm) RuO$_2$ on TiO$_2$, Fig.~\ref{fig:ZFCFC}(c), is used as a control measurement for comparison to the heterostructure samples. Here, the susceptibility data shown was fit to a traditional Curie-Weiss law with a constant background susceptibility $\chi_0$. The $\chi_0$ was temperature independent, which is emblematic for Pauli paramagnetic materials.\cite{ryden1970magnetic, Fletcher1968} After the background was subtracted, the inverse susceptibility $\chi^{-1}$ was plotted as seen Fig.~\ref{fig:ZFCFC}(f). The Curie-Weiss law was fit between approximately 26 K to 110 K since the data behaved fairly linearly in that region, where $\theta_{CW} \approx$ -8 K and $C \approx$ 2.06x10$^{-6}$ m$^{3}$K/mol. Below 26 K, we observe a very weak 1/T feature, which is the inverse behavior to the heterostructures within a similar temperature range. These measurements and fitting generally support the original claim that RuO$_{2}$ is a Pauli paramagnet with the presence of an additional low temperature feature that deviates from the Curie-Weiss law.\cite{ryden1970magnetic} However, 4$d$ transition metals that display metallic behavior, like RuO$_{2}$, are known to generally deviate from the linear Curie-Weiss law behavior, which may account for the low temperature 1/T behavior seen in the pure RuO$_2$.\cite{mugiraneza2022tutorial}
\begin{figure*}[]
    \centering
    \includegraphics[width =0.95\textwidth]{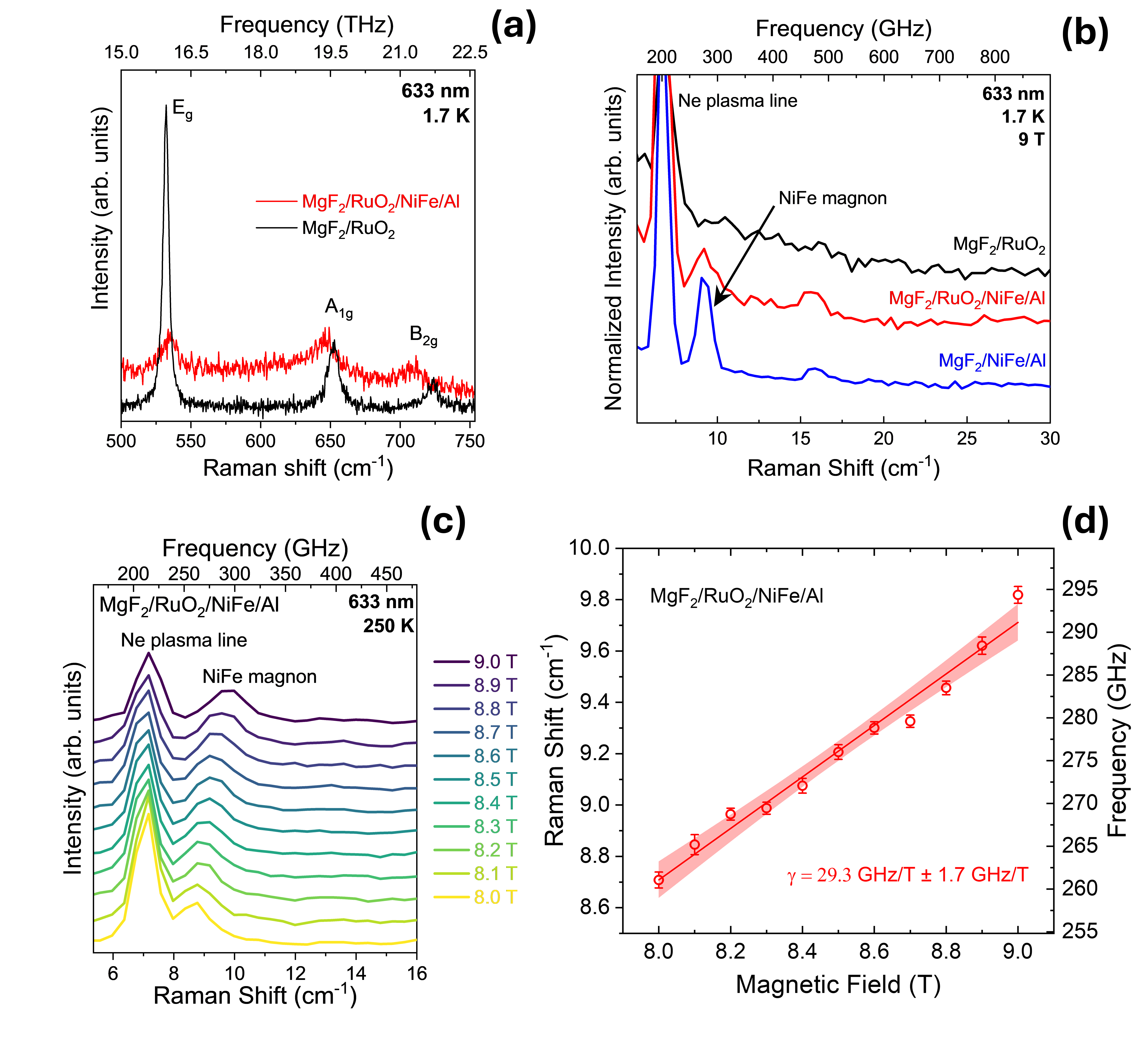}
    \caption{Raman and Magneto-Raman measurement of (100) RuO$_2$. Raman spectra collected at 1.7 K from MgF$_{2}$/(100) RuO$_{2}$, and MgF$_{2}$/(100) RuO$_{2}$/NiFe/Al showing phonon modes and assigned symmetries. Fitted frequencies are shown in Table ~S3. Magneto-Raman spectra in the low-frequency region (GHz range) collected at 1.7 K with 9 T magnetic field applied in the plane i.e. Voigt configuration. The NiFe-capped RuO$_2$ sample shows a peak at 9.4 cm$^{-1}$ that is consistent with the NiFe magnon mode observed in the MgF$_2$/NiFe/Al control sample (b). Field dependent Raman measurements between 8 T and 9 T of MgF$_2$/(100) RuO$_2$/NiFe/Al sample at 250 K (c), and determination of gyromagnetic ratio at 250 K of the NiFe magnon (d).}
    \label{fig:raman}
\end{figure*}

To investigate possible long range magnetic ordering, interfacial effects, and predictions of magnons in RuO$_2$, \cite{garcia2024magnon, vsmejkal2023chiral} temperature-dependent, magneto-Raman measurements were performed. Three samples were considered: (1) (100) RuO$_2$ orientation grown on MgF$_2$ with nominal thickness of 100 nm, capped with 10 nm of NiFe and 5 nm Al, (2) a reference sample (100) RuO$_2$ orientation grown on MgF$_2$ also with nominal thickness of 100 nm and a 5 nm Al cap, and (3) a reference sample of 10 nm NiFe grown on MgF2 capped with Al. Importantly, the heterostructure (1) of RuO$_2$ and NiFe showed the emergence of exchange bias at a similar temperature to the primary heterostructures discussed in the magnetometry measurements and these data are plotted in the supplemental material in Fig. S3 and S4.  The phonon behavior at zero magnetic field of the (100) RuO$_2$ film, both with and without the 10 nm NiFe capping layer were collected at 1.7 K using 632.82 nm excitation from a HeNe laser. Fig.~\ref{fig:raman}(a) shows the Raman spectra of the two RuO$_2$ films on MgF$_2$ substrates; with and without NiFe cap. (Of note, the TiO$_2$ substrates have phonons in the spectral window that interfered with those of RuO$_2$ and therefore where not used for Raman.) 
\par
The prominent Raman modes of RuO$_2$(100) observed are E$_g$ (532.12$^{-1}$ $\pm$ 0.01 cm$^{-1}$), A$_{1g}$, (652.18$^{-1}$ $\pm$ 0.09 cm$^{-1}$) and B$_{2g}$ (732.12$^{-1}$ $\pm$ 0.25 cm$^{-1}$) for the RuO2/MgF2 in agreement with previous works.\cite{mar1995characterization, meng2003raman, jevtic2006noise} See Table S3 for literature and theory comparison. While the spectra of both samples feature the characteristic phonon modes of RuO$_2$, the NiFe cap significantly attenuates the RuO$_2$ Raman signal, as expected, resulting in a reduced signal-to-noise ratio. The shifting of the peaks with and without the cap likely results from strain. 
\par
Next, we applied a 9 T magnetic field parallel to the sample surface and measured the Raman response of the three structures. As shown in Fig.~\ref{fig:raman}(b), both the MgF$_2$/RuO$_2$/NiFe and MgF$_2$/NiFe samples exhibit a low-frequency peak $\approx$ 277 GHz (9.2 cm $^{-1}$). In contrast, this feature is absent in the MgF$_2$/RuO$_2$ sample, indicating that the observed peak is associated with the presence of the NiFe layer. Extending the spectral range to 11.8 THz shows that no additional modes are observed for all three samples (Fig. S5). This frequency range covers the normal range for magnon modes typically observed in AFM materials \cite{mccreary2020quasi}, suggesting a lack of long range magnetic ordering in RuO$_2$. Temperature dependent Raman of the MgF$_2$/RuO$_2$/NiFe sample at a fixed applied field of 9 T is shown in Fig. S6 and reveals minimal shift in the mode with temperature, even around the region where the spin-like transition was observed in the ZFC data ($\approx$ 31 K) of the other heterostructure samples. To further confirm the observed mode in the heterostructure belongs to NiFe, field-dependent measurements at 250 K were performed from 8 T to 9 T, showing a shift to higher wavenumber with increasing field as shown in Fig.~\ref{fig:raman}(c). Assuming the mode is a magnon due to NiFe, the calculation of g$\mu_\mathrm{B}$ at 250 K was 29.3 GHz/T $\pm$1.7 GHz/T (g = 2.09 $\pm$ 0.12) shown in Fig.~\ref{fig:raman}(d).  The accepted value of Ni$_{80}$Fe$_{20}$’s magnon mode from Brillouin light scattering measurements is 29.4 GHz/T at 300 K.\cite{liu2005magnetic} To further confirm that the magnon originates from the NiFe layer and that the underlying RuO$_2$ does not measurably influence its response, we compared the magnetic field dependence of the magnon at 1.7 K with that of the control MgF$_2$/NiFe sample (Figure S7). Both samples exhibit comparable g$\mu_\mathrm{B}$ values of 30.8 GHz/T $\pm$ 4.3 GHz/T and 30.5 GHz/T $\pm$ 1.0 GHz/T for MgF$_2$/RuO$_2$/NiFe and MgF$_2$/NiFe respectively.  The lack of intrinsic magnon modes in RuO$_2$ or any new mode not associated with NiFe in the heterostructure suggests the observed magnetic phenomenon is coming from a local ordering effect at the interface (spin, diffusion, or a combination of both).

\begin{figure}[t]            
  \centering
  \includegraphics[width=\columnwidth]{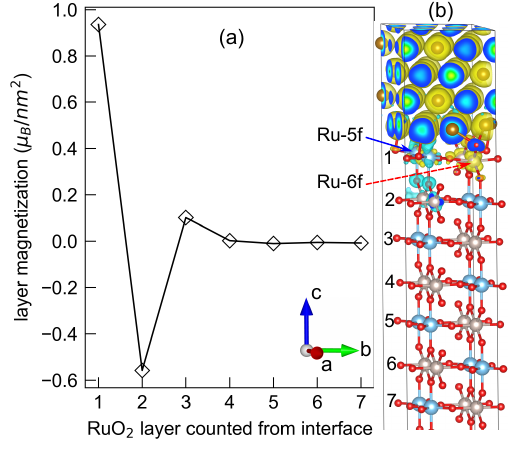}
  \caption{%
    (a) Magnetization ($\mu_\mathrm{B}$/nm$^{2}$) as a function of RuO$_2$ layer in
    contact with Fe.  
    (b) Atomic structure of the RuO$_2$(110)/Fe(001) interface with spin-density
    isosurfaces. Yellow and cyan colors show spin-up and spin-down densities, respectively. Two Ru sublattices are represented by different colors; O and Fe are in red and gold colors, respectively.}
  \label{fig:dft}
\end{figure}

In order to further understand the interfacial properties of RuO$_2$/ferromagnetic heterostructures, we performed DFT calculations, specifically for system consisting of RuO$_2$(110) and bcc Fe(001). In our previous work, we investigated the freestanding (110) surface of RuO$_2$ and found that although bulk RuO$_2$ has a nonmagnetic ground state, the RuO$_2$(110) surface exhibits spontaneous magnetization and spin-polarized surface states due to the breaking of local symmetry. \cite{ho2025symmetrybreakinginducedsurfacemagnetization,torun2013role} Similar to the freestanding RuO$_2$(110) surface, the RuO$_2$(110)/Fe(001) bilayer exhibits spin-polarized behavior at the interface. Fig. \ref{fig:dft}(a) displays the magnetization (in $\mu_\mathrm{B}$/nm$^{2}$) as a function of layer within the RuO$_2$(110) film that is in contact with Fe. Each ``layer" of RuO$_2$ consists of an equal number of Ru atoms belonging to two structurally equivalent Ru sublattices of the bulk with the attached O atoms. These Ru sublattices are 6-fold coordinated everywhere except at the surface where we have 5-fold coordinated Ru atom (Ru-5f) and a 6-fold coordinated Ru atom (Ru-6f) due to to surface termination. However, to facilitate our analysis, we denote the two Ru sublattices as Ru-5f and Ru-6f consistent with their structural positions: those being on the same sublattice with the Ru-5f atom at the surface will be denoted by Ru-5f and similarly for Ru-6f. Fig.~\ref{fig:dft}(b) displays the spin density isosurfaces of the heterostructure. Similar to the freestanding RuO$_2$(110)\cite{ho2025symmetrybreakinginducedsurfacemagnetization}, we observe significant spin-polarization on the surface of RuO$_2$(110) (layer 1) for the heterostructure with Fe. In the freestanding structure, the magnetization per layer quickly converges towards 0 $\mu_\mathrm{B}$/nm$^{2}$ for subsurface layers ($>$ layer 1). In contrast to freestanding structure, we observe a significant antiparallel magnetization in the second layer of Ru below the surface for the RuO$_2$(110)/Fe bilayer (see Fig. \ref{fig:dft}). The antiparallel magnetization gives rise to a quasi-AFM-like coupling between the surface and the subsurface layers when RuO$_2$(110) is in contact with Fe(001). In addition, the atomic magnetic moment of Fe at the interface increases from $\approx$ 2.2 $\mu_\mathrm{B}$ to $\approx$ 2.6 $\mu_\mathrm{B}$, signifying strong interaction between Fe and RuO$_2$. 
Fig.~S8 shows a more in-depth perspective of Fig. \ref{fig:dft}, showing the individual contributions of Ru and O to the layer-dependent magnetization. From this, we also see that the Ru-5f and Ru-6f atoms are spin-polarized in opposite directions throughout the heterostructure, and the sign of these moments changes between the surface (layer 1) and penetrating layers within the RuO$_2$ (layer 2 and so on), and most strikingly for the Ru-6f atoms. In addition, it is evident that the presence of the Fe layer significantly impacts the magnetization of O atoms on the surface and the second layer below the surface. Fig \ref{fig:dft} and Fig.~S8 emphasize the possible spin disorder at the interface between RuO$_2$(110) and Fe(001) which can give rise to AFM-like effects as seen from the anti-parallel coupling between Ru-5f and Ru-6f atoms within the same layer and the anti-parallel coupling between the surface and the first penetrating layer (layer 2) within the RuO$_2$. 

To gain further insight regarding the interfacial interaction, we computed the charge density difference between the heterostructure RuO$_2$(110)/Fe(001) and its constituents, i.e., the isolated RuO$_2$(110) and Fe(001) parts (see Fig.~S9(a)). As seen in Fig.~S9(a)), charge is slightly transferred from the Fe layer to RuO$_2$, but mostly localized at the interface which can be due to screening in this metallic system. Fig.~S9(b)) depicts the planar and macroscopic averaged potential profiles across the film normal direction, showing a higher potential for Fe, which further confirms the charge transfer from the Fe layer to the RuO$_2$ layer. As expected, the charge transfer and significant change in magnetization of the surface O and the Fe at the interface result in changes to the bonding environment between Fe and O. From Fig.~\ref{fig:dft}, Fig.~S8, and Fig.~S9, we observe bonding between surface O atoms and the Fe layer, with O approaching the Fe layer. This interaction could signify a possible diffusion of O atoms into the ferromagnetic layer.     

\section*{Discussion}

Here, we propose two possible hypotheses that might explain the origin of observed exchange bias, spin-like transition feature at 31 K, and the lack of observed magnon modes by Raman spectroscopy. The first hypothesis is that there is an intrinsic spin disorder at the RuO$_{2}$/FM interface and formation of Fe-O bonds as indicated by DFT calculations. The second mechanism is that the effects seen in the magnetometry are from diffusion/intermixing at the interface, including the formation of a discrete disordered Fe-O (passed the interface bonding) or diffusion of Fe/Ni ions into the non-magnetic rutile structure, creating a spin-glass in contact with the FM. Similar materials have been studied in the context of dilute magnetic semiconductors with interstitial Fe in a matrix, leading to spin-glass behavior,\cite{ma2015structure,heiman1992laser,lakshmi2004spin,furdyna1986diluted}, where exchange bias can be observed. Additionally, interfacial AFM phase have been observed to form due to diffusion and cause exchange bias in topological insulator ferromagnetic heterostructures. \cite{bhattacharjee2022effects, bhattacharjee2022topological, will2023antiferromagnetic} However, we note that the formation of the stoichiometric AFM FeO phase is unlikely due to the following reasons. Previous studies have shown that significant diffusion can occur at the interface of a rock salt structure. Deposition of Co on ZnO has been shown to form an AFM interfacial phase\cite{10.1063/5.0209098}; this is similar to the case of Fe grown on MgO where the magnetic atoms can substitute into the substrate crystal structure.\cite{WANG2020147501,nnano.2013.894} However, it is improbable that this type of diffusion occurred based on our observed magnetic properties. The Fe-Ru-O phase diagram was examined, and there are unlikely to be any possible AFM ternary phases that could form at the biasing temperature of 400 K.\cite{10.1002/pssa.2211080130,fujimori1969magnetic,10.1063/1.4812323}  Furthermore, the only energetically favorable crystal structures that could be induced at the interface leading to the AFM ordering for both samples is FeO and Hematite $\alpha-$Fe$_2$O$_3$, which have N\'eel temperatures of $198$ K \cite{10.1063/5.0082729} and $>900$ K\cite{10.1063/5.0094868}, respectively. Since the EB is only seen at low temperatures and the spin transition is observed at about 31 K, the formation of a stoichiometric interfacial AFM phase is unlikely. All of these scenarios could result in observed exchange bias, magnetic frustration as observed in exponentially decaying coercivity/EB, and a low-temperature spin transition. Regardless, the lack of observed magnons is consistent with the purely interfacial nature of these effects and supported by other studies, which suggest a lack of long-range magnetic ordering. \cite{PhysRevLett.132.166702, npjSpintronics2024}

\section*{Conclusion}

The possibility of a new type of magnetic ordering leading to large band splitting is an exciting prospect for the future of spintronics. However, the magnetic ordering in RuO$_2$ has been debated over the past year, with several studies pointing toward RuO$_2$ not being magnetic, or exhibiting only defect-based magnetism. In our study, we examined two independently grown RuO$_2$ heterostructures with different ferromagnetic layers which exhibit similar exchange bias fields with temperature and a spin-like transitions at about 31 K. The transition is only clearly seen when a ferromagnet is layered on the top of the RuO$_2$ thin film, suggesting an interface effect. The observation of exponential decay of both EB and $H_c$ supports magnetic frustration, with magneto-Raman revealing a lack of long-range magnetic ordering in the RuO$_2$. The findings primarily support an interfacial spin pinning effect, possibly caused by interface diffusion or intrinsic spin disorder, as shown by DFT.

\section*{Methods}

Thin films of RuO$_2$ were grown using two deposition chambers via reactive magnetron sputtering. The first set was grown on MgF$_2$ and TiO$_2$ comprising of RuO$_2$/NiFe/Al. With the sample of primary focus being grown on (110) MgF$_2$. Another sample grown in a different chamber comprised of RuO$_2$/Fe/Ru and was deposited on (110) TiO$_2$. EB results for the additional samples are reported in the supplemental material.\cite{SM} The heterostructure of RuO$_2$/NiFe/Al on (110) MgF$_2$ was prepared through a magnetron sputtering process in a vacuum chamber with a base pressure maintained below 3x$10^{-8}$ Torr. The substrates were preheated to 450$^{\circ}$C and the RuO$_2$ layer was deposited by sputtering a 2" diameter pure Ru (99.95\%, Angstrom Engineering) target using DCX 750-4 Source at 200 W in an Ar/O$_2$ gas mixture (7 sccm each). The total pressure was maintained at 15 mTorr. After depositing this layer, the substrates were left to cool naturally to ambient temperature overnight. Subsequently, in-situ deposition of NiFe layer was performed by applying 100 W RF power across 2" diameter NiFe (nominally composition of 81 at. \% Ni 19 at. \% Fe) target under an Ar pressure of 5 mTorr with a flow rate of 30 sccm. Finally, an in-situ deposition of a capping layer of Al was done by applying 200 W DC power across 2" diameter pure Al target under an Ar pressure of 3 mTorr with a flow rate of 30 sccm. During the deposition of all layers, the substrate holder was rotated at 20 RPM to ensure uniform film thickness. The heterostructure comprising RuO$_2$/Fe/Ru on (110) TiO$_2$ was prepared through a reactive DC magnetron sputtering process in an AJA ATC Orion deposition system.\cite{10.1021/acs.cgd.4c00271} For this sample, the substrate was first heated to 450 $^\circ$C in 15 mTorr of Ar (7.5 sccm) and O$_2$ (7.5 sccm) and allowed to equilibrate for 15 minutes to remove surface organics. After, the background pressure was reduced to 4 mTorr, and growth of the 35 nm-thick RuO$_2$ layer was conducted through the application of 30 W DC across a 2" diameter pure Ru (99.5\%, ACI alloys) target using a DCXS-750-4 Multiple Sputter Source. After RuO$_2$ deposition, the substrate was allowed to cool to ambient temperature. Then, a 10 nm-thick Fe layer was grown through the application of 40 W DC across a 1.5" diameter pure Fe (99.5\%, ACI alloys) target using an Advanced Energy MDX-1k power supply in 3 mTorr of Ar (15 sccm). After, a 5 nm-thick Ru capping layer was grown at ambient temperature using the the same target and power supply as the RuO$_2$ layer, but with 3 mTorr of Ar (15 sccm) as the background pressure and with 40 W DC.
\par
The structure of the thin films was characterized by X-ray diffraction (XRD) using a Bruker D8 Discover system equipped with a rotating anode Cu K$\mathit{\alpha}$ source and a Dectris Eiger 2R 500K area detector. High resolution XRD (HRXRD) intensity pattern and reciprocal space map (RSM) measurements were collected using a Ge (400) 2-bounce monochromator, 0.4 mm incident slit, and 0.5 mm collimator, whereas X-ray reflectivity (XRR) measurements were made using a Goebel mirror and only a 0.5 mm collimator. 
\par
Magnetometry measurements were performed on two different Quantum Design (QD) systems. The magnetometry measurements for the data presented in the supporting information and  the MgF$_{2}$/NiFe/Al reference sample were performed in a 14 T Quantum Design PPMS Dynacool platform using the VSM option in a temperature range from 2 K to 400 K. The samples were directly glued to the quartz VSM sample holder using rubber cement. Before starting measurement, samples were heated to 400 K and an external magnetic field (1 T) was applied along the easy axis for 1 hour. Then, the samples were subsequently cooled down to approximately 2K (base temperature) in presence of 1 T field, and M vs. H loop measurements were taken from 2K to 300K on warming. All other data presented in the main text with the exception of the EB and coercivity data points for the MgF$_{2}$/NiFe/Al sample in Fig. \ref{fig:Temp_EB}(c,d) were performed on a QD MPMS3 SQUID magnetometer system using an identical field cooling procedure. In this instrument, samples were affixed to a quartz holder with a thin piece of kapton tape before loading into the chamber and field cooling ahead of measurement. M vs. T measurements were conducted with an applied field of 5 mT and 10 mT for the MgF$_{2}$ and TiO$_{2}$-based heterostructures, respectively, and for M vs. H loop measurements  out to 100 mT to 150 mT were collected on warming.
\par

In-plane magnetic field and temperature dependent Raman measurements in the 180$^{o}$ backscattering configuration were performed using an attoDRY 2100 cryostat (attocube Inc.) with a (Horiba JY T64000, 1800 groove/mm grating) spectrometer, which is coupled to a liquid-nitrogen cooled CCD detector, using a 632.82 nm excitation from a HeNe gas laser. In this Voigt configuration, with the sample’s surface parallel to the applied field, the excitation laser is redirected with a mirror and focused with a lens into the sample surface with a spot size of approximately 10 µm. Samples were cooled to the base temperature 1.7 K without the application of an applied field, followed by both field and temperature dependent measurements on warming. 
\par
To gain insight into the effects of RuO$_2$ and Fe at the interface, we performed DFT \cite{PhysRev.136.B864,PhysRev.140.A1133} calculations using the projector augmented wave (PAW) method\cite{PhysRevB.54.11169,PhysRevB.59.1758}, the Vienna {\it Ab initio} Simulation Package (VASP), and the Perdew-Burke-Ernzerhof (PBE) \cite{PhysRevLett.77.3865} functional. We simulated a heterostructure composed of rutile RuO$_2$(110) and bcc Fe(001). The heterostructure was built using a superlattice approach consisting of thirteen layers of RuO$_2$ along [110] crystallographic direction and nine layers of Fe along [001]. The optimized thirteen-layer structure of RuO$_2$(110) was obtained from our previous study\cite{ho2025symmetrybreakinginducedsurfacemagnetization}, which focused on the freestanding surface. To understand the effect of ferromagnetic layers on RuO$_2$, we fixed the in-plane lattice constants of RuO$_2$, while unavoidable strain was applied to the Fe layer due to lattice mismatch between the two materials. Using the InterMat \cite{D4DD00031E} package, we were able to construct an affordable structure with 219 atoms. To optimize atomic configurations, we fixed the seven most inner layers of RuO$_2$ and three most inner layers of the Fe layer, leaving three layers of RuO$_2$ and three layers of Fe at the interface relaxed until the Hellmann-Feynman force on each relaxing atom is smaller than 0.010 eV/\AA. Gamma-centered k-mesh of 10$\times$8$\times$1 was employed to sample the Brillouin zone, and a cutoff energy of 600 eV was used to expand the wavefunctions. Valence electronic configurations of Ru, O, Fe were $4p^64d^75s^1$, $2s^22p^4$, $3d^74s^1$, respectively.

\section*{Disclaimer}

Certain commercial equipment, instruments, software, or materials are identified in this paper in order to specify the experimental procedure adequately. Such identifications are not intended to imply recommendation or endorsement by the National Institute of Standards and Technology (NIST), nor it is intended to imply that the materials or equipment identified are necessarily the best available for the purpose.

\begin{acknowledgements}
This research was sponsored by NSF DMR- 2316664, NSF through the University of Delaware Materials Research Science and Engineering Center, DMR-2011824. The authors acknowledge the use of the Materials Growth Facility (MGF) at the University of Delaware, supported by NSF through the University of Delaware Materials Research Science and Engineering Center, DMR-2011824. It also used resources from the National Energy Research Scientific Computing Center (NERSC), a Department of Energy Office of Science User Facility, using the NERSC award BES-ERCAP 0034471 (m5002).
Research at the United States Naval Academy was supported by the Kinnear Fellowship and Office of Naval Research under Contract No. N0001423WX02132 and continued ONR support. F.M.A. and B.D. would like to acknowledge the support of Mr. Peter Morrison and the Office of Naval Research under Contract No.~N00014-21-WX-01248. This work was supported by the Office of Naval Research 6.1 Base Funding at the U.S. Naval Research Laboratory. R.T. would like to acknowledge the NIST/National Research Council Postdoctoral Research Associateship Program for funding.
\end{acknowledgements}
\newpage
\label{References}
\bibliographystyle{apsrev4-1}
\bibliography{bib}
\end{document}


\author{Frank M. Abel$^\dagger$}
\affiliation{Physics Department, United States Naval Academy, Annapolis, MD 21402, USA}
\affiliation{National Institute of Standards and Technology, Gaithersburg, MD 20899, USA}

\author{Subhash Bhatt$^\dagger$}
\affiliation{Department of Physics and Astronomy, University of Delaware, Newark, DE 19716, USA}

\author{Shelby S. Fields$^\dagger$}
\affiliation{Materials Science and Technology Division, U.S. Naval Research Laboratory, Washington, District of Columbia 20375, USA}

\author{Vinay Sharma}
\affiliation{Laboratory for Physical Sciences, College Park, MD 20740, USA}

\author{Dai Q. Ho}
\affiliation{Department of Materials Science and Engineering, University of Delaware, Newark, Delaware 19716, USA}
\affiliation{Faculty of Natural Sciences, Quy Nhon University, Quy Nhon 55113, Vietnam}

\author{Daniel Wines}
\affiliation{National Institute of Standards and Technology, Gaithersburg, MD 20899, USA}

\author{D. Quang To}
\affiliation{Department of Materials Science and Engineering, University of Delaware, Newark, Delaware 19716, USA}

\author{Joseph C. Prestigiacomo}
\affiliation{Materials Science and Technology Division, U.S. Naval Research Laboratory, Washington, District of Columbia 20375, USA}

\author{Tehseen Adel}
\affiliation{National Institute of Standards and Technology, Gaithersburg, MD 20899, USA}
\affiliation{Department of Physical Sciences, University of Findlay, Findlay, OH 45840}

\author{Riccardo Torsi}
\affiliation{National Institute of Standards and Technology, Gaithersburg, MD 20899, USA}

\author{Maria F. Munoz}
\affiliation{National Institute of Standards and Technology, Gaithersburg, MD 20899, USA}

\author{David T. Plouff}
\affiliation{Department of Physics and Astronomy, University of Delaware, Newark, DE 19716, USA}

\author{Xinhao Wang}
\affiliation{Department of Physics and Astronomy, University of Delaware, Newark, DE 19716, USA}

\author{Brian Donovan}
\affiliation{Physics Department, United States Naval Academy, Annapolis, MD 21402, USA}

\author{Don Heiman}
\affiliation{Department of Physics, Northeastern University, Boston, MA 02115, USA}
\affiliation{Plasma Science and Fusion Center, MIT, Cambridge, MA 02139, USA}

\author{Gregory M. Stephen}
\affiliation{Laboratory for Physical Sciences, College Park, MD 20740, USA}

\author{Adam L. Friedman}
\affiliation{Laboratory for Physical Sciences, College Park, MD 20740, USA}

\author{Garnett W. Bryant}
\affiliation{National Institute of Standards and Technology, Gaithersburg, MD 20899, USA}
\affiliation{Joint Quantum Institute, University of Maryland, College Park, Maryland 20742, USA}

\author{Anderson Janotti}
\affiliation{Department of Materials Science and Engineering, University of Delaware, Newark, Delaware 19716, USA}

\author{Michelle E. Jamer}
\affiliation{Physics Department, United States Naval Academy, Annapolis, MD 21402, USA}

\author{Angela R. Hight Walker}
\affiliation{National Institute of Standards and Technology, Gaithersburg, MD 20899, USA}

\author{John Q. Xiao}
\affiliation{Department of Physics and Astronomy, University of Delaware, Newark, DE 19716, USA}

\author{Steven P. Bennett}
\affiliation{Materials Science and Technology Division, U.S. Naval Research Laboratory, Washington, District of Columbia 20375, USA}

\title{Supplementary Material for\\ ``Probing Magnetic Properties of RuO$_{2}$ Heterostructures Through the Ferromagnetic Layer"}
\maketitle

\section{Crystallographic Information}

\begin{figure*}[!h]
    \centering
    \includegraphics[width = 0.8\textwidth]{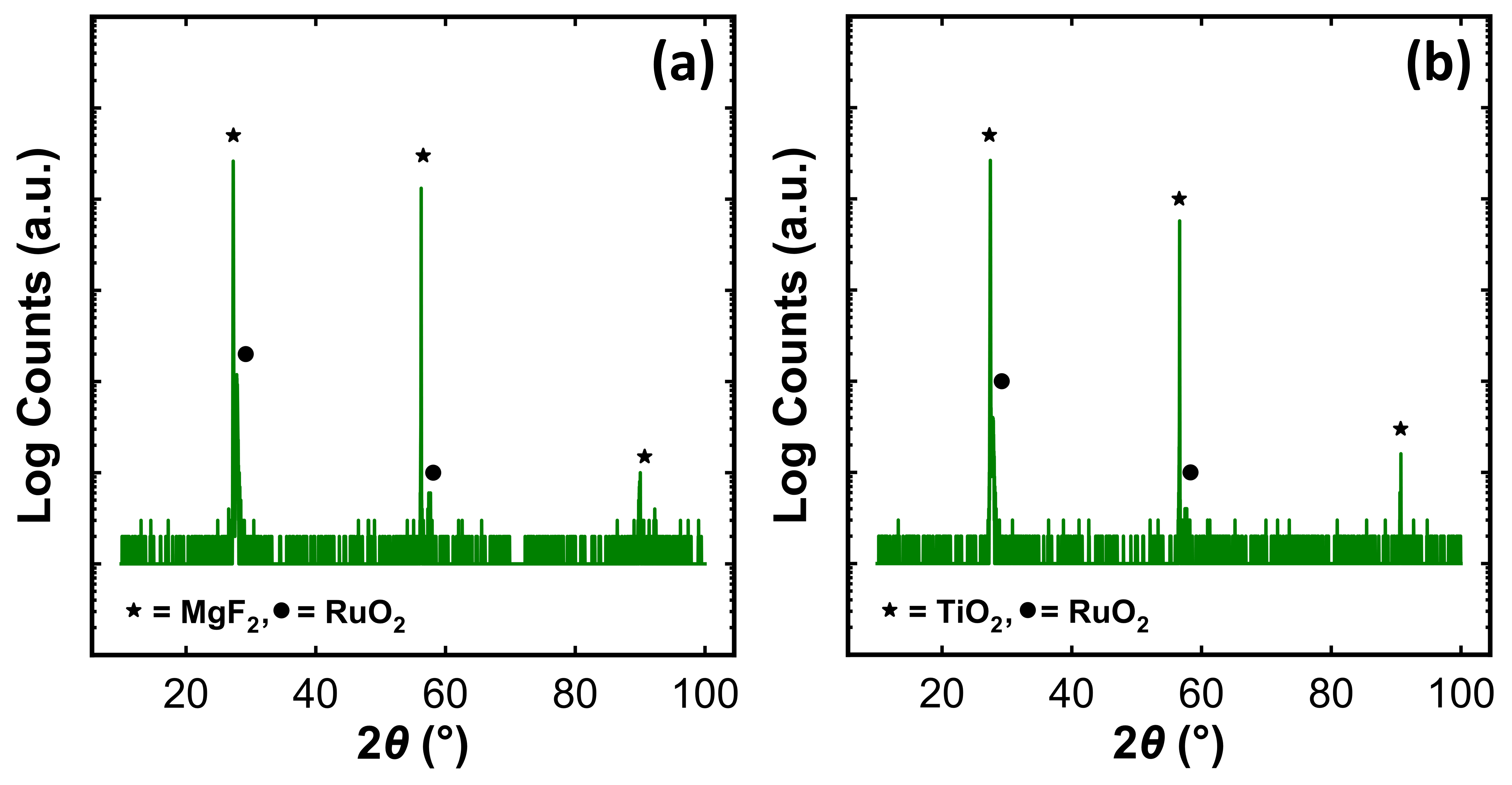}
     \caption{XRD of RuO$_2$/NiFe/Al on MgF$_{2}$ (a) and RuO$_2$/Fe/Ru on TiO$_{2}$ (b).}
    \label{fig:xrd}
\end{figure*}

X-ray diffraction from the samples are reported in Fig. \ref{fig:xrd}(a) and (b) for the MgF$_2$ and TiO$_2$ samples, respectively. As noted in the main text, the crystal structure is in the (110) direction for both samples and the substrate peaks are noted by stars and the dots note the structure peaks. The film indicated epitaxial growth for both substrates.

\begin{figure*}[]
    \centering
    \includegraphics[width = 0.8\textwidth]{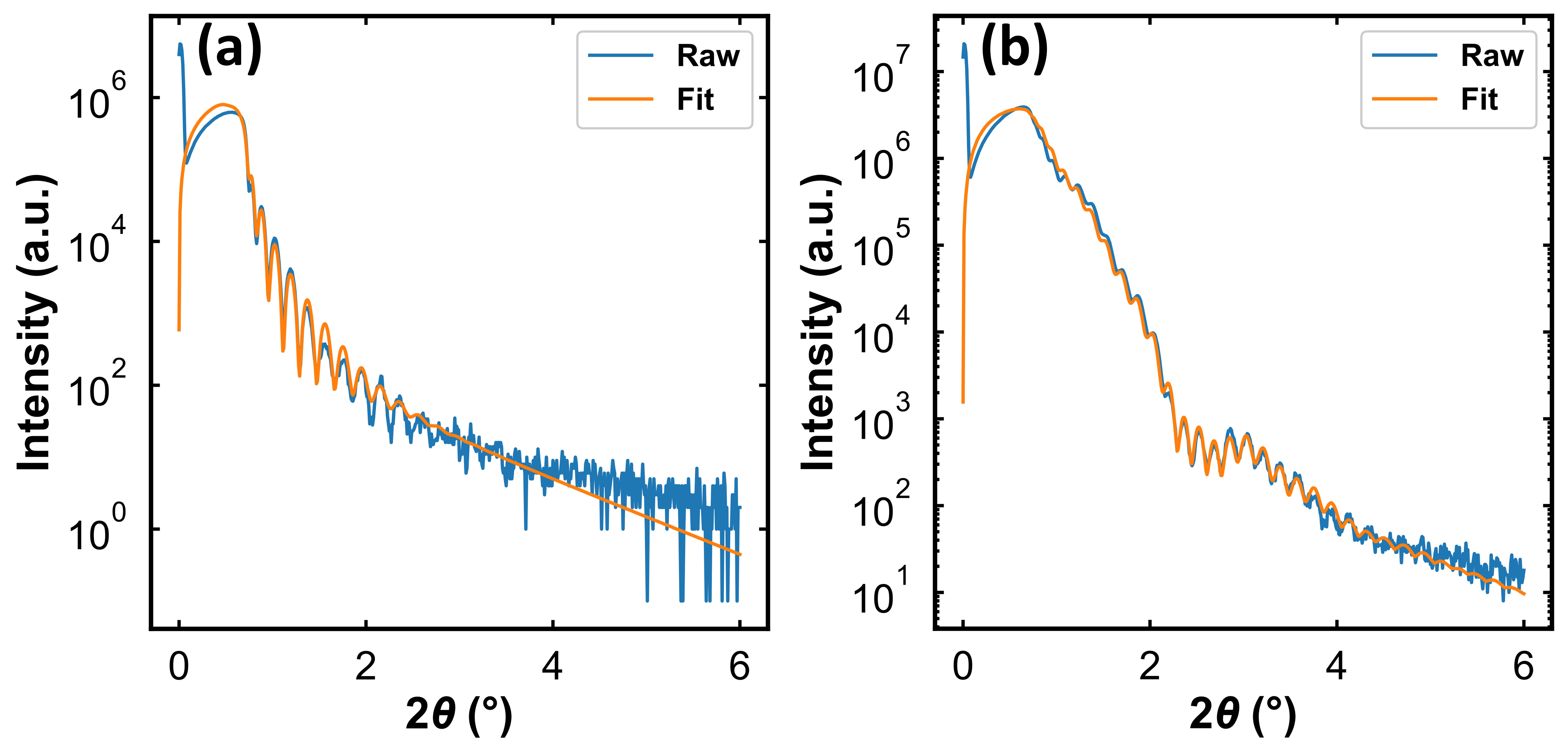}
     \caption{XRR and fit of RuO$_2$/NiFe/Al on MgF$_{2}$ (a) and RuO$_2$/Fe/Ru on TiO$_{2}$ (b).}
    \label{fig:xrr}
\end{figure*}

The thicknesses of the RuO$_2$ and ferromagnetic layer reported in the main text were found by using X-ray reflectivity (XRR). The XRR data for both samples can be seen in Fig. \ref{fig:xrr} for the MgF$_2$ and TiO$_2$ samples in (a) and (b), respectively. The subsequent fits are illustrated in the figure, and the values of thicknesses for the layers are noted in Table \ref{tab:XRR_fit}.

\begin{table}[!ht]
    \centering
    \begin{tabular}{|c|c|c|c|}
    \hline
     &  Layer  & MgF$_2$ &TiO$_2$ \\
\hline
 \multirow{3}{*}{Thickness (nm)}    & RuO$_2$ & 32.3 & 35.8 \\
    & FM &  8.8 & 9.5 \\
    & Capping Layer & 4.3 & 3.8 \\
    \hline
 \multirow{3}{*}{Roughness (nm)}    & RuO$_2$ & 10.8 & 0.5 \\
    & FM &  1.2 & 0.6 \\
    & Capping Layer & 2.3 & 0.9 \\
    \hline    
    \end{tabular}
    \caption{The parameters for XRR from the fit in Fig. \ref{fig:xrr}. The table shows the comparable nature of the samples chosen for the study including the thicknesses. The roughness noted in the table indicates that the RuO$_2$ is rougher in the MgF$_2$ substrate.}
    \label{tab:XRR_fit}
\end{table}

\section{Model Fitting Equations and Parameters}

The exponential decay was fit to the following equations:

\begin{equation}
    H_c = H_0 e^{-\frac{T}{T_0}} + H_1 T + H_2
\end{equation}
\begin{equation}
    H_{EB} = H_{EB,0} e^{-\frac{T}{T_0}} + H_{EB,1} T + H_{EB,2}
\end{equation}

\noindent where $H_0$ in Eq. 1. corresponds to the coercivity at T = 0 K, and T$_0$ is the temperature below which has an enhanced coercivity from the pinning causing the EB.

\begin{table}[!ht]
    \centering
    \begin{tabular}{|c|c|c|c|c|c|}
    \hline
     Sample & Field Value & $T_0$ & $H_0$ & $H_1$ & $H_2$ \\
     \hline
    \multirow{2}{*}{MgF$_2$/RuO$_2$/NiFe/Al} & $H_c$ & 14.4 & 54 & -0.078 & 39.9\\
     & $|$$H_{EB}$$|$ & 4.3 & 35 & 0 & 1.3\\
    \hline 
    \multirow{2}{*}{TiO$_2$/RuO$_2$/Fe/Ru} & $H_c$ & 25 & 22 & -0.006 & 2.2\\
        & $|$$H_{EB}$$|$ & 5.6 & 29 & 0.002 & 1.4\\
    \hline    
    \end{tabular}
    \caption{Fitting parameters for the $H_{EB}$ and $H_c$}
    \label{tab:ebfits}
\end{table}

\section{Exchange Bias in Additional Samples}

\begin{figure*}[h!]
    \centering
    \includegraphics[width = 1\textwidth]{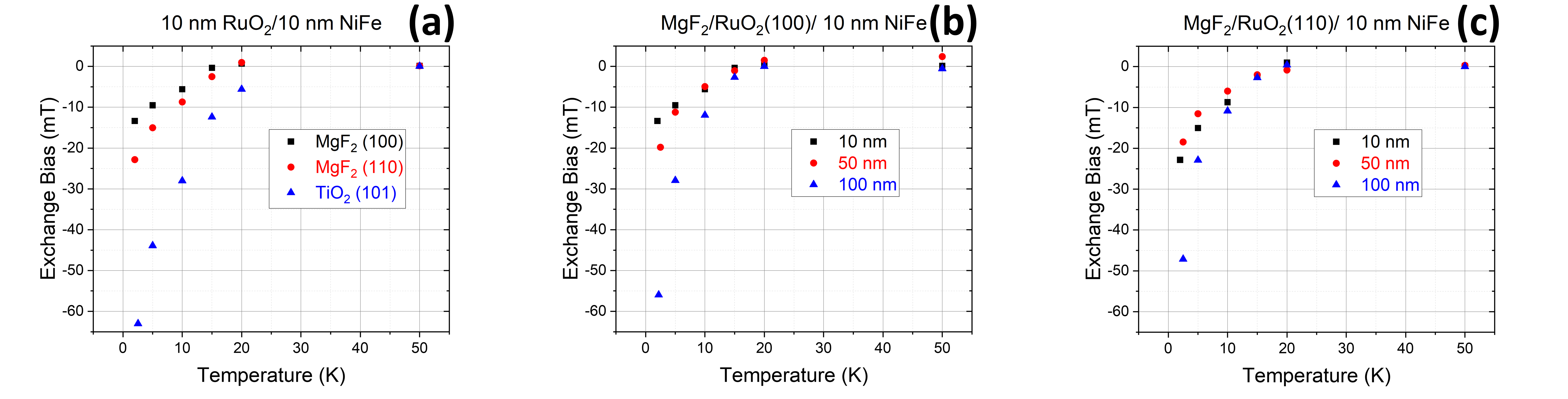}
     \caption{Summary of exchange bias (EB) for samples with a fixed thickness and different orientations (a), different thicknesses of RuO$_{2}$ with (100) orientations (b), and thicknesses of RuO$_{2}$ with (110) orientations. Thickness values are based on deposition rate.}
    \label{fig:EB_all}
\end{figure*}

\begin{figure*}[]
    \centering
    \includegraphics[width = 1\textwidth]{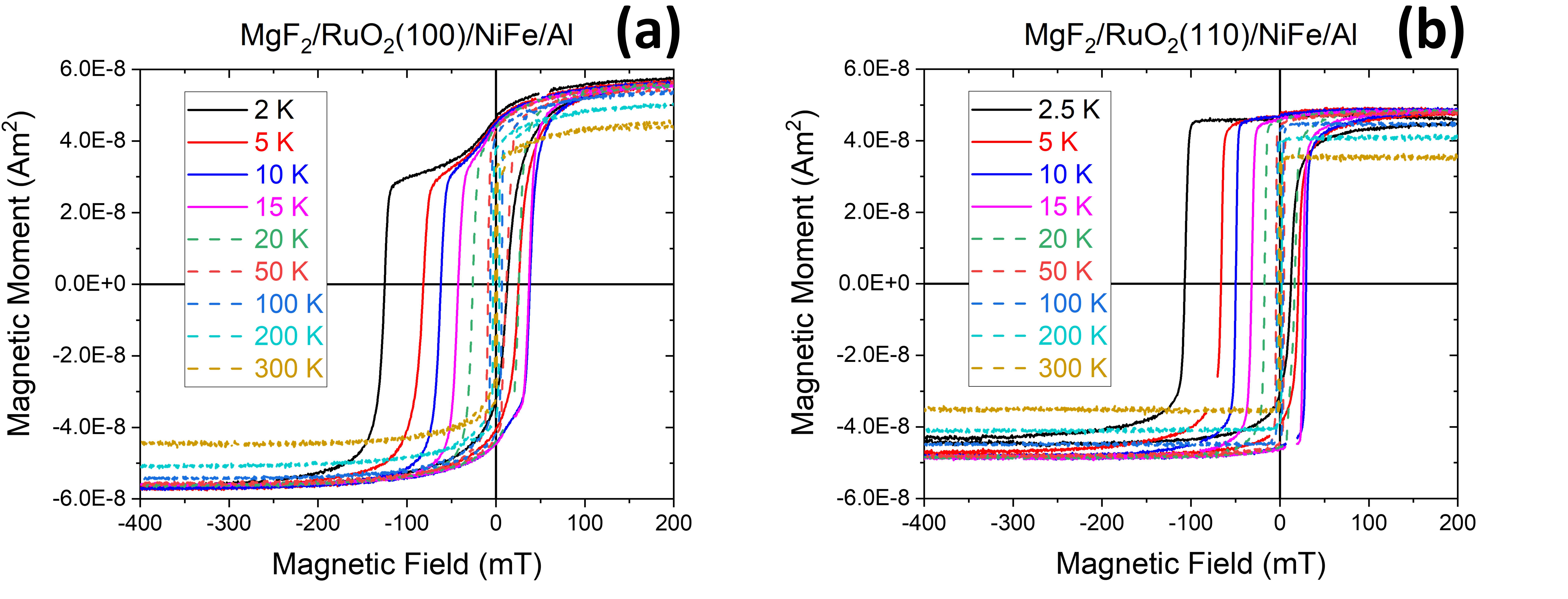}
     \caption{Hysteresis from about 2 K to 300 K of MgF$_{2}$/RuO$_{2}$/NiFe/Al with 100 nm (100) oriented RuO$_{2}$ used for Raman study (a), and of MgF$_{2}$/RuO$_{2}$/NiFe/Al with 100 nm (110) oriented RuO$_{2}$, which shows similar shift in both orientations.}
    \label{fig:MvsHRaman}
\end{figure*}

The exchange bias effect was observed in other orientations of the RuO$_2$ thin films as illustrated in Fig. \ref{fig:EB_all}. Each sample has a 10 nm NiFe layer to generate the EB effect. The thicknesses reported in these figures are determined by deposition rate. Fig. \ref{fig:MvsHRaman} shows the hysteresis loops from about 2 K to 300 K for the two thickest RuO$_{2}$ samples considered, one which was used for the magneto-Raman study. Each sample was biased by heating to 400 K, then applying a $+1$ T field for one hour, then cooled to 2 K.

\section{Additional Raman Spectroscopy Data}

\begin{table}[h]
\begin{adjustbox}{width=\columnwidth,center}
\begin{tabular}{|l|ll|ll|ll|ll|}
\hline
Sample & Single Crystal \cite{mar1995characterization}                      &  & MgF$_2$/RuO$_2$                       &  & MgF$_2$/RuO$_2$/NiFe/Al                      &  & Theory \cite{choudhary2020joint, wines2023recent}                      &  \\ \hline
Raman Peak & \multicolumn{1}{l|}{Freq. (cm$^{-1}$)} & FWHM (cm$^{-1}$) & \multicolumn{1}{l|}{Freq. (cm$^{-1}$)} & FWHM (cm$^{-1}$) & \multicolumn{1}{l|}{Freq. (cm$^{-1}$)} & FWHM (cm$^{-1}$) & \multicolumn{1}{l|}{Symmetry} & DFT (cm$^{-1}$) \\ \hline
1 & \multicolumn{1}{l|}{528} & 11 & \multicolumn{1}{l|}{532.12 $\pm$ 0.01} & 4.75 $\pm$ 0.04 & \multicolumn{1}{l|}{534.33 $\pm$ 0.30} & 15.11 $\pm$ 0.95 & \multicolumn{1}{l|}{E$_{g}$} & 495.1 \\ \hline
2 & \multicolumn{1}{l|}{646} & 15 & \multicolumn{1}{l|}{652.18 $\pm$ 0.09} & 9.40 $\pm$ 0.26 & \multicolumn{1}{l|}{644.75±0.40} & 25.52±1.28 & \multicolumn{1}{l|}{A$_{1g}$} & 610.8 \\ \hline
3 & \multicolumn{1}{l|}{716} & 15 & \multicolumn{1}{l|}{723.12 $\pm$ 0.25} & 10.09 $\pm$ 0.72 & \multicolumn{1}{l|}{709.03 $\pm$ 0.62} & 13.32 $\pm$ 1.90 & \multicolumn{1}{l|}{B$_{2g}$} & 695.1 \\ \hline
\end{tabular}
\end{adjustbox}

\label{tab:Phonon_modes}
\caption{Comparison of the fitted RuO$_{2}$ Raman modes with theory and literature values.}
\end{table}

\begin{figure*}[]
    \centering
    \includegraphics[width = 1\textwidth]{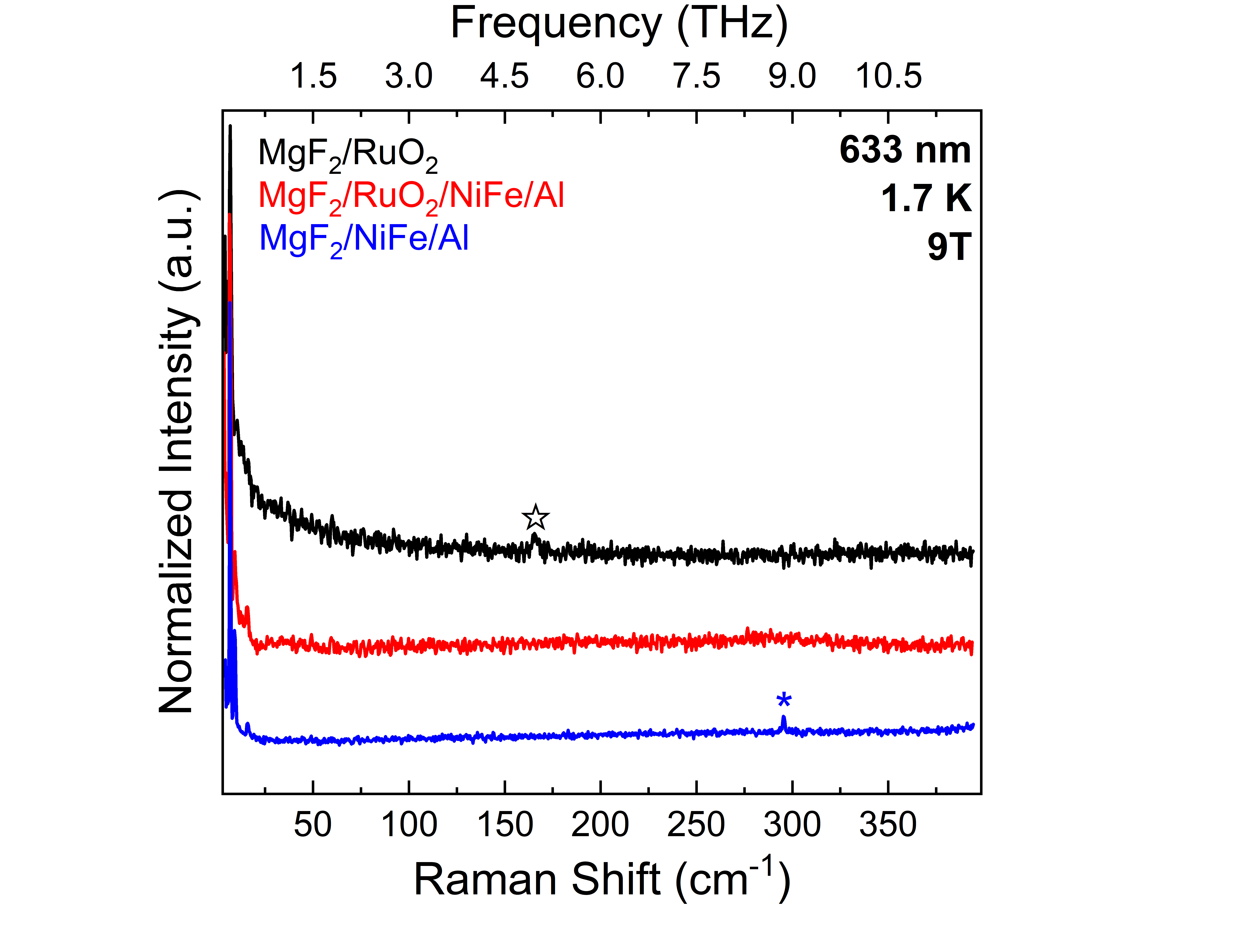}
     \caption{Raman measurements at 9 T applied magnetic field, collected at 1.7 K of MgF$_{2}$/(100) RuO$_{2}$, MgF$_{2}$/(100) RuO$_{2}$/NiFe/Al, and MgF$_2$/NiFe/Al samples showing no additional modes beyond the NiFe magnon peak at $\approx$ 9.4 cm$^{-1}$. The starred peak in black is the weak scattering B1g phonon mode in RuO2. The B1g peak is not visible in the NiFe-capped RuO$_2$ sample since the signal is significantly attenuated by the capped as evidenced in Fig. 5a. The asterisk in blue is a peak coming from the MgF$_{2}$ substrate.}
    \label{fig:S5_Raman}
\end{figure*}

\begin{figure*}[]
    \centering
    \includegraphics[width = 1\textwidth]{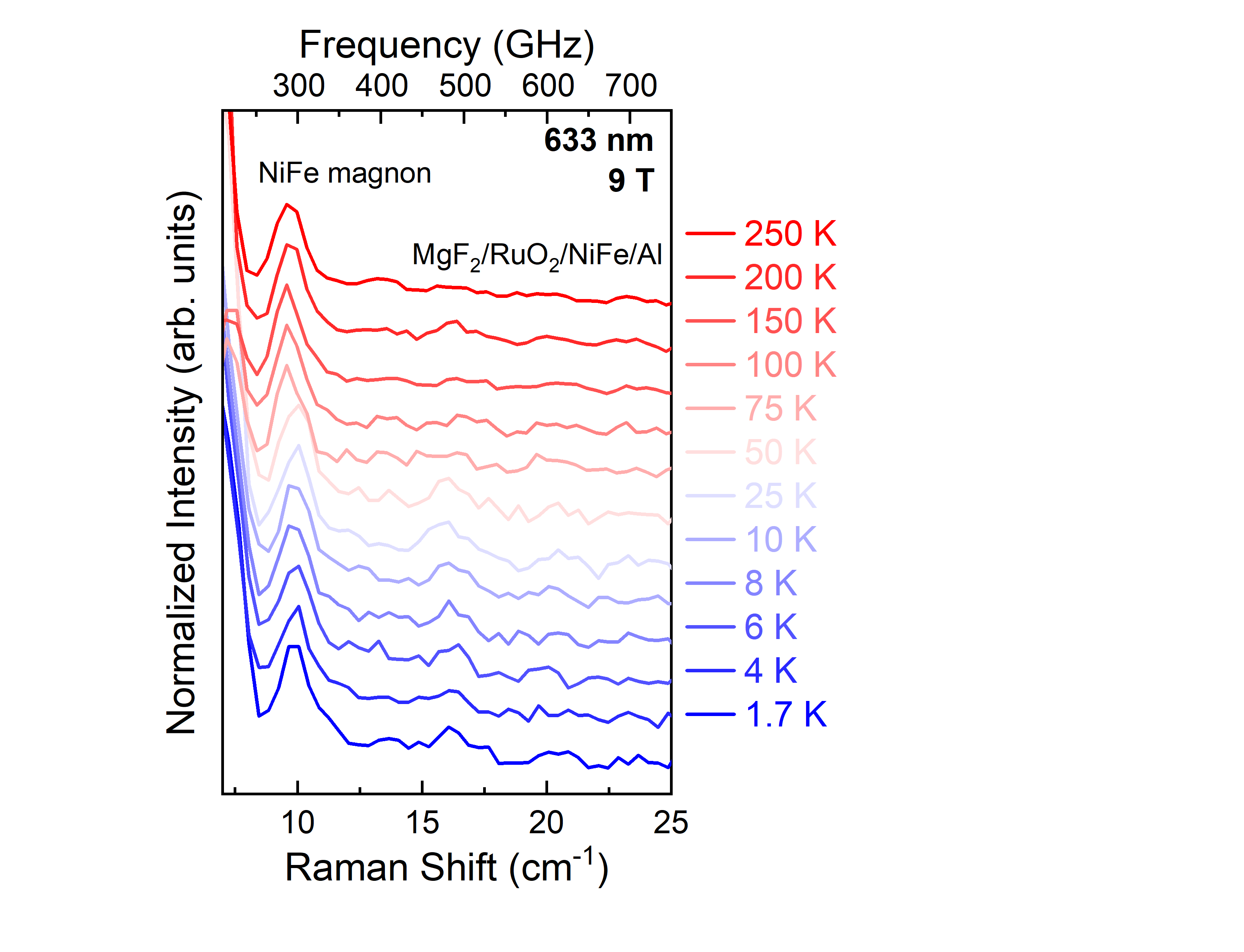}
     \caption{Temperature dependent Raman measurements at 9 T applied magnetic field of the NiFe magnon of the MgF$_{2}$/(100) RuO$_{2}$/NiFe/Al sample. As expected, the plot shows no measurable shift in magnon mode in this temperature range since the NiFe Curie temperature is $\approx$ 750 K.\cite{zhang2019understanding}}
    \label{fig:S6_Raman}
\end{figure*}

\begin{figure*}[]
    \centering
    \includegraphics[width = 1\textwidth]{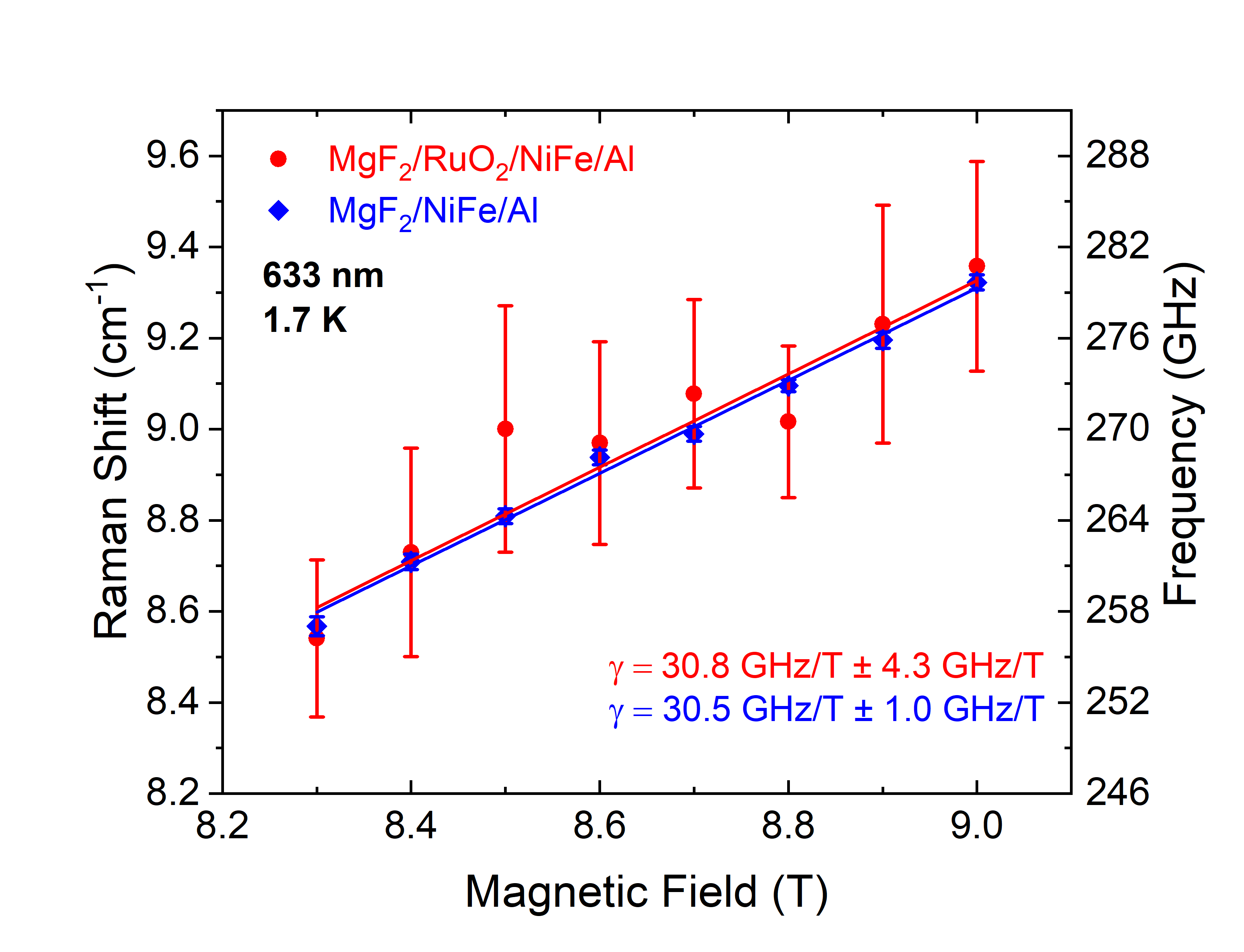}
     \caption{Determination of the gyromagnetic ratio at 1.7 K of the NiFe magnon of the MgF$_{2}$/(100) RuO$_{2}$/NiFe/Al (red) and MgF$_{2}$/NiFe/Al (blue) samples. The Raman shift of the magnon mode as a function of field is plotted from 8.3 T to 9.0 T where the magnon peak is clearly separated from the laser plasma line. }
    \label{fig:S7_Raman}
\end{figure*}

\section{Density Functional Theory Results}

\begin{figure*}[h]
    \centering
    \includegraphics[width = 1\textwidth]{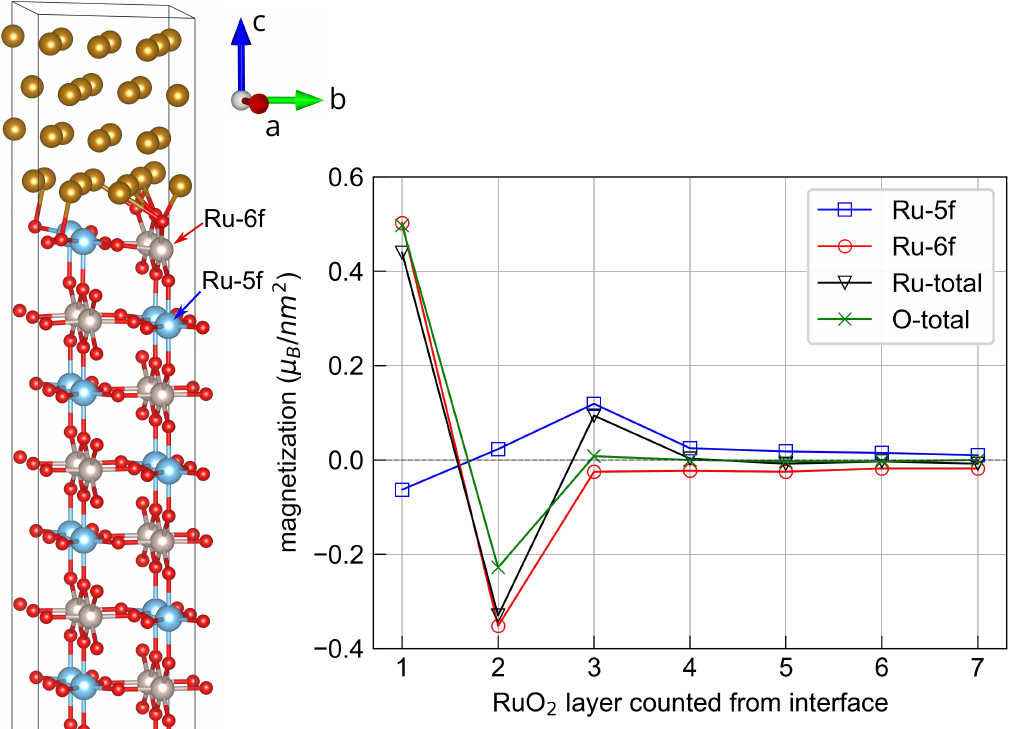}
     \caption{Density functional theory (DFT) magnetization (in $\mu_B$/nm$^{2}$) as a function of layer within the RuO$_2$ that is in contact with Fe for the 5-fold coordinated Ru (Ru-5f), 6-fold coordinated Ru (Ru-6f), total Ru per layer and total O per layer. The atomic structure of the RuO$_2$(110)/Fe(001) bilayer is given as reference (silver atoms are Ru-6f and blue atoms are Ru-5f).}
    \label{fig:dfts1}
\end{figure*}

\begin{figure*}[]
    \centering
    \includegraphics[width = 1\textwidth]{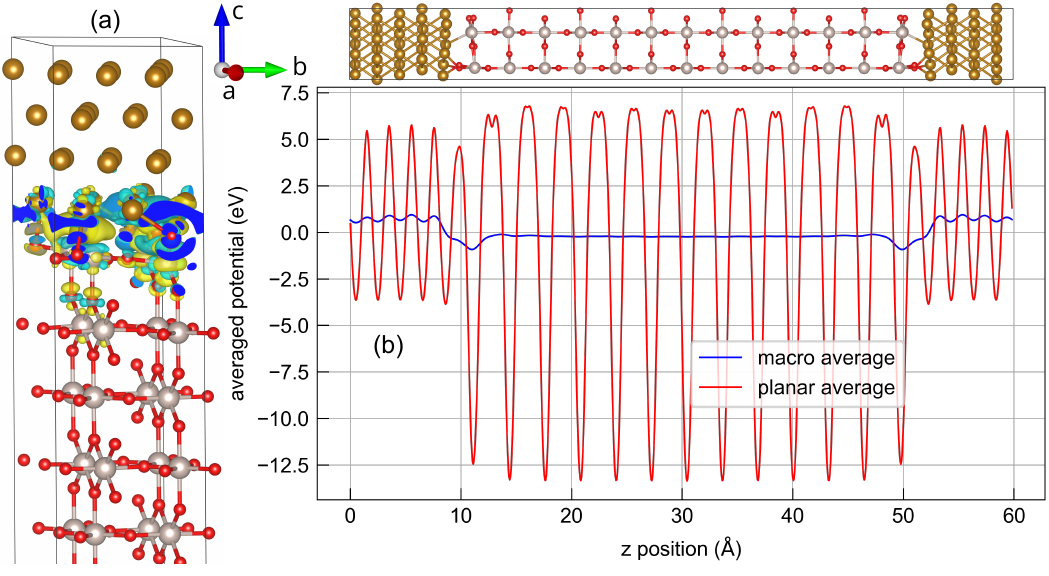}
     \caption{Depicts the DFT charge density difference between the RuO$_2$(110) and Fe(001) layers in the hetrostructure. Yellow signifies charge accumulation while cyan signifies depletion (a). Depicts the planar and macroscopic averaged potential profiles across the z direction (with the atomic structure of the heterostructure for reference) (b).}
    \label{fig:dfts2}
\end{figure*}

\clearpage
\label{References}
\bibliographystyle{apsrev4-1}
\bibliography{bib}

\section*{Disclaimer}

Certain commercial materials, equipment, instruments, and software are referenced in this manuscript to clearly describe the experimental and computational procedures. These references to not imply an endorsement or recommendation by NIST, or does it mean these are the best suited for the described purposes. The authors declare no competing financial interest.